\documentclass[showpacs,twocolumn]{revtex4-1}
\usepackage{epstopdf}
\usepackage{color}
\usepackage{bm}
\usepackage{graphicx}
\usepackage{amsmath}
\usepackage{amsfonts}
\usepackage{amssymb}
\usepackage{bezier} 
\usepackage{color}

\newcommand{\mib}{\bm}
 
\def\beq{\begin{equation}}
\def\eeq{\end{equation}}
\def\beq{\begin{equation}}
\def\eeq{\end{equation}}
\def\bea{\begin{eqnarray}} 
\def\eea{\end{eqnarray}} 
\def\ssr#1{{\scriptscriptstyle{\rm #1}}}
\def\frac#1#2{{\textstyle{#1 \over #2}}}
\def\bnabla{\boldsymbol{\nabla}}
\def\half{\frac{1}{2}} 
\def\pz{\partial}
\def\eps{\epsilon}
\mathchardef\sPi="7105
\mathchardef\sSigma="7106
\mathchardef\sPhi="7108
\mathchardef\sLambda="7103
\mathchardef\sOmega="710A
\mathchardef\sTheta="7102
\def\Tra{\mathop{\textsf{Tr}}}
\def\ST{\textsf{T}}
\def\SP{\textsf{P}}
\def\SD{\textsf{D}}
\def\SB{\textsf{B}}
\def\Sb{\textsf{b}}
\def\SA{\textsf{A}} 
\def\nd{^{\vphantom\dagger}}
\def\Bpi{{\mib\pi}}
\def\BPhi{{\mib\sPhi}}

\def\Bx{{\mib x}}
\def\Beta{{\mib\eta}}
\def\Bq{{\mib q}}
\def\Bp{{\mib p}}
\def\Bk{{\mib k}}
\def\BS{{\mib S}}
\def\BE{{\mib E}} 
\def\BPhi{{\mib \Phi}} 
\def\CO{{\cal O}}
\def\CL{{\cal L}}
\def\RA{{\rm A}}
\def\RB{{\rm B}}
\def\RC{{\rm C}}

\def\RE{{\rm E}}

\def\nhat{{\hat{\mib n}}}
\def\stil{{\widetilde\sigma}}

\def\Imp{\textsf{Im}\,}
\def\bvph{\vphantom{\sum_N^N}}
\def\ie{{\it i.e.\/}}
\def\SH{S\nd_0}
\def\SAn{S\nd_{\textsf A}}
\def\SC{S\nd_{\textsf C}}
 
\begin{document}

\title{Visibility of the Amplitude (Higgs) Mode in Condensed Matter}
\author{Daniel Podolsky,$^1$ Assa Auerbach,$^{1,2}$ and Daniel P. Arovas$^3$}
\affiliation{ $^1$Physics Department, Technion, 32000 Haifa,
Israel\\
$^2$Department of Physics, Stanford University, Stanford CA 94306, USA\\
$^3$Department of Physics, University of California at San Diego, La Jolla, California 92093, USA.}


\date{November 23, 2011}

\begin{abstract}
The amplitude mode is a ubiquitous collective excitation in  condensed-matter systems with broken continuous symmetry.
It is expected in antiferromagnets, short coherence length superconductors, charge density waves, and lattice Bose condensates.
Its detection is a valuable test of  the corresponding field theory, and its mass gap measures the proximity to a quantum critical point.
However, since the amplitude mode can decay into low-energy Goldstone modes, its experimental visibility has been questioned. 
Here we show that the visibility depends on the symmetry of the measured susceptibility.  
The longitudinal susceptibility diverges at low frequency as $\chi_{\sigma\sigma}\sim i\omega^{-1}$ $(d=2)$ or $\log(1/|\omega|)$ $(d=3)$,  
which can completely obscure the amplitude  peak. In contrast,  the scalar susceptibility is suppressed by four extra powers of frequency, exposing the
amplitude peak throughout the ordered phase. We discuss experimental setups for measuring the scalar susceptibility.
The conductivity of the $O(2)$ theory  (relativistic superfluid) is a scalar response and therefore exhibits 
suppressed absorption below   the Higgs mass threshold,  $\sigma \sim \omega^{2d+1}$.  In layered, short coherence length superconductors,
(relevant e.g. to cuprates) this threshold is raised by the
interlayer plasma frequency.
\end{abstract}


\maketitle
\section{Introduction}
A fundamental consequence of spontaneous (continuous) symmetry breaking (SSB) of $N$-component order parameters
is the emergence of collective order parameter oscillations: Goldstone modes and a massive amplitude (Higgs) mode.  
Examples in condensed matter systems are plentiful: for example, Heisenberg and $XY$ spin systems,  superconductors,
cold atom condensates in optical lattices and incommensurate charge density waves.\cite{relativistic}  In particle theory, the Higgs boson \cite{Higgs} is   
modeled by the amplitude mode of a gauged bosonic condensate.

\begin{figure}[!t]
\begin{center}
\includegraphics[width=7cm,angle=0]{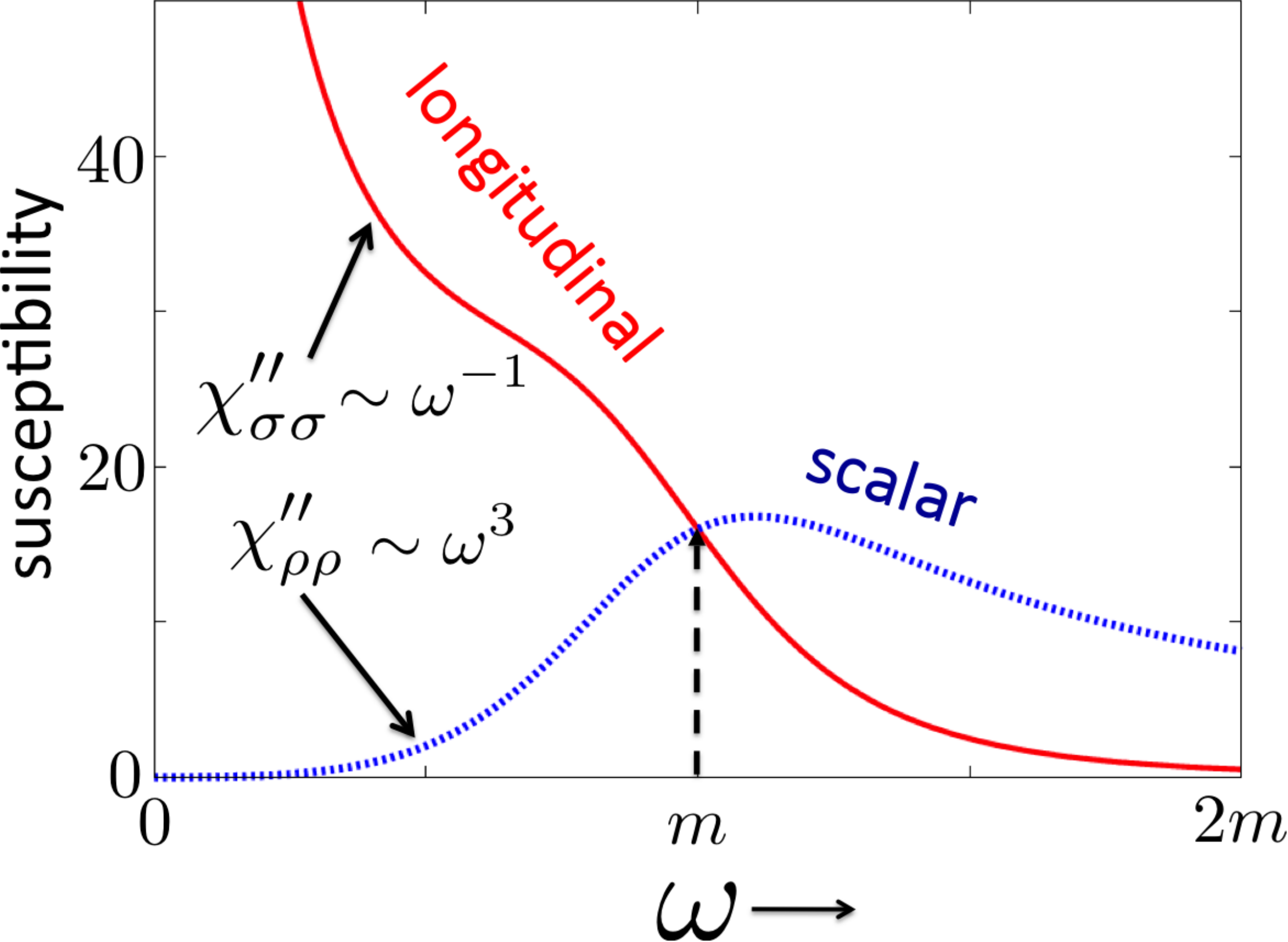}
\caption{{\bf Visibility of the amplitude mode.} The longitudinal and scalar susceptibilities  of the  two-dimensional $O(N)$ model at zero temperature are plotted
in the large $N$ limit. $m$ is the renormalized amplitude mode mass.
The coupling constant is $g=0.84\, g_{\rm c}^\infty $, which is within the broken symmetry phase. 
Note that dissipation into Goldstone modes has very different effects on the low-frequency behavior of the two  susceptibilities and on
the visibility of the amplitude mode around $\omega=m$.}
\label{chifig} 
\end{center}
\end{figure}

While long-wavelength Goldstone modes are sharp excitations in the broken symmetry phase,  the amplitude mode is long-lived only at the classical (weak coupling) level.  Quantum corrections allow for its decay into pairs of  Goldstone modes.    Previous authors \cite{Subir,Zwerger}  have
therefore questioned the experimental visibility of the amplitude mode.  Indeed, in ordered phases of two and three dimensions, the longitudinal susceptibility exhibits an infrared singularity, which can broaden the  amplitude mode peak into an undetectable shoulder, as in Fig.~\ref{chifig}.

{\em Is the amplitude mode therefore overdamped?  } This paper shows that, surprisingly, it is not.  Rather, the infrared singularity is a property of the chosen
susceptibility.  The amplitude mode produces, in fact, a pronounced peak in the scalar (rotationally invariant) susceptibility, as depicted in
Fig.~\ref{chifig}.  Similarly, for the $O(2)$ model (relativistic superfluids), the mode produces a well-defined pseudogap in the optical conductivity,  as previously noted in Kubo formula computations.\cite{LA-PRB}  Here we find at two-loop order an additional weak absorption tail below the Higgs mass threshold,  $\sigma \sim \omega^{2d+1}$.  

The two types of susceptibility are associated with different ways of parametrizing the fluctuations of the order parameter ${\bf \Phi}$.  Within the ordered phase, the order parameter gets an expectation value $\left|\langle{\bf \Phi}\rangle\right|=\Phi_0$, and fluctuations in its value can be written as
\begin{eqnarray}
\begin{split}
{\bf \Phi}&=\left(\Phi_0+\sigma,\Bpi\right)\\
  &=\Phi_0 \left(1+\sqrt{N}\rho\right) \nhat.
\end{split} 
\label{parametrizations}
\end{eqnarray}
In the first parametrization, the fields $\sigma$ and $\Bpi$ describe, in turn, the longitudinal and transverse fluctuations relative to the ordering direction.  In the second parametrization, $\rho$ is the fluctuation in the magnitude of the order parameter, while $\nhat$ is the ordering direction.    We find that the scalar susceptibility, associated with $\rho$, can display a sharp peak at the Higgs mass, even in cases where the longitudinal susceptibility, associated with $\sigma$, does not. 

The difference between longitudinal and amplitude fluctuations can be understood heuristically by considering oscillations of a particle near the
minimum of the Mexican hat potential shown in Fig.~\ref{fig:MH}.   The longitudinal component (in the broken symmetry direction) loses coherence
rapidly as the particle oscillates and also meanders around the rim.  
In contrast, amplitude oscillations, that is, fluctuations in the radial distance $\rho$, are much longer lived since their interaction with long-wavelength Goldstone fluctuations is suppressed by two derivatives.  Thus, the infrared
singularity in the longitudinal response is due to   ``contamination''  of Goldstone modes in the response function and not due to overdamping of the amplitude mode 
itself.\cite{Popov,Dupuis}  This explanation is made precise by calculating  the scalar susceptibility of the relativistic
$O(N)$  field theory  in the broken symmetry phase.

This paper is organized as follows: In Sec.~\ref{sec:fieldTheory}, we introduce the $O(N)$ field theory on which we base our analysis, define longitudinal and scalar susceptibilities of the order parameter, and show how to define conductivity in $O(N)$ models.  In Sec.~\ref{sec:weakCoupling} we compute in the weak coupling limit the susceptibilities to one-loop order and the optical conductivity up to two-loop order. We show that a cancellation of self-energy and vertex diagrams suppress the  sub-gap  spectra  of the scalar
susceptibility and the $O(2)$ conductivity by four powers of frequency relative to non symmetric response functions.  In Sec.~\ref{sec:largeN} we compute the $O(N)$ susceptibilities in the large $N$ limit and find a suppression by the same four powers of frequency in the scalar susceptibility relative to the longitudinal susceptibility.  We conclude in Sec.~\ref{sec:experiments} by proposing various scalar probes for the amplitude mode  in cold atoms, magnetic systems, and layered  superconductors with long-range Coulomb interactions.  In the cuprates, for example, we suggest that the mid-infrared  spectral weight in the optical conductivity might  be related to the amplitude mode of tightly bound Cooper pairs.\cite{BasovTimusk}  This is followed by appendixes which provide detailed derivations of some of the technical results discussed in the main text.

\section{$O(N)$ Field Theory}
\label{sec:fieldTheory}
The relativistic  $O(N)$ field theory describes long-wavelength correlations of condensed matter systems with
$N$ real parameter fields $\BPhi$, with global $O(N)$ symmetry and two time derivatives in the Lagrangian.\cite{QPT}  For example, $O(3)$ theory describes
unfrustrated Heisenberg antiferromagnets,\cite{Haldane} and $O(2)$ theory describes strongly interacting lattice bosons at commensurate~\cite{Fisher}
and half-commensurate\cite{LA-PRB} fillings.  The role of the speed of light is played by a spin wave or sound velocity.
The Euclidean time action reads as
\beq 
S={1\over 2g}\int_\sLambda \! d^{d+1}\!x\left[ (\pz_\mu \BPhi)^2+{m_0^2\over 4N}\left(|\BPhi|^2-N\right)^2\right].
\label{eq:action}
\eeq 
Here $\BPhi$ is dimensionless, the momenta integrals are cut off by the ultraviolet wave vector $\sLambda$, and the bare mass $m\nd_0$ is given by the microscopic scale of the system, of order $\sLambda$.   

Note that Eq.~(\ref{eq:action}) can be regarded as a coarse-grained version of the $O(N)$ non linear $\sigma$ model.  Very deep inside the ordered phase, the $\sigma$ model does not allow for amplitude fluctuations; that is, the amplitude excitations are infinitely gapped.  More generally, the model can be coarse-grained to a soft-spin model, as treated here, and the amplitude excitations have a finite energy.

\begin{figure}[!t]
\begin{center}
\includegraphics[width=7cm,angle=0]{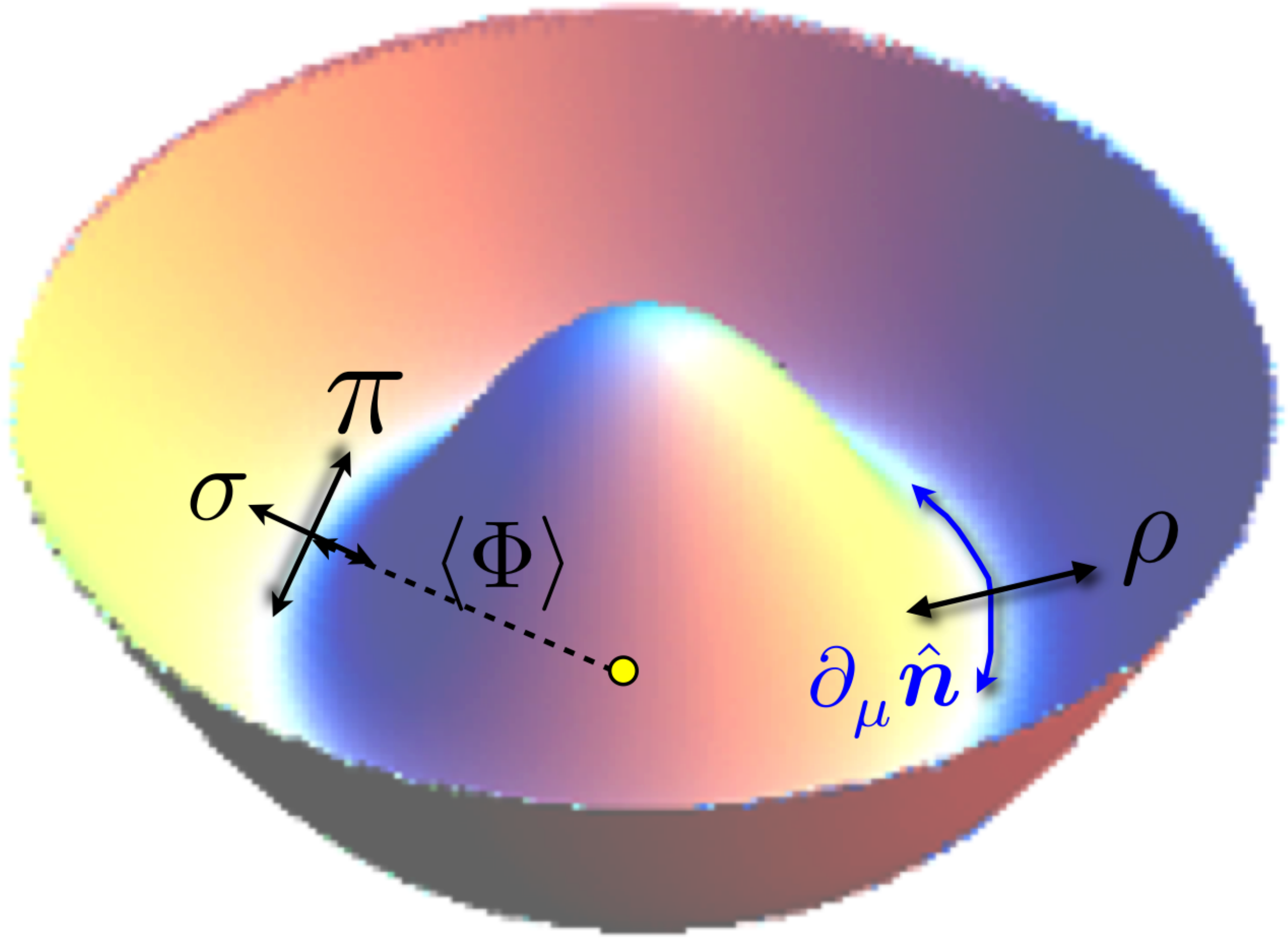}
\caption{{\bf Fluctuations in the Mexican hat potential.} $\langle\BPhi\rangle$ is the order parameter.  The massive Higgs and massless Goldstone modes can be represented either in terms of longitudinal ($\sigma$) and transverse ($\Bpi$) degrees of freedom, respectively, or in terms of scalar ($\rho$) and direction ($\partial_\mu \nhat$) degrees of freedom.}
\label{fig:MH}
\end{center}
\end{figure}
For  $d\ge 2$, there is an ordered SSB phase at weak coupling $g < g_{\rm c}(\sLambda)$. Deep in the ordered phase ($g\ll g_{\rm c}$) the order parameter is $|\langle \BPhi\rangle|\simeq\sqrt{N}$.  Quantum fluctuations reduce the order parameter to $ r (g,\sLambda)\sqrt{N}$, while the amplitude mode mass is reduced from its bare value $m\nd_0$ to a renormalized value $m(g,\Lambda)$ until they  both vanish at the quantum critical point
$g_{\rm c}$~\cite{QPT}.   For Eq.~(\ref{eq:action})  to describe the amplitude mode, the cutoff must be large enough to satisfy
$\sLambda\gg m$, which holds particularly well in the vicinity of a  quantum critical point.

Fluctuations in the broken symmetry phase can be parametrized by
\beq 
\BPhi=( r\sqrt{N}+\sigma \, , \, \Bpi)\ ,
\label{expand}
\eeq
where $r$ is a constant. $\sigma$ is the longitudinal ({\em not scalar!})  fluctuation and
$\Bpi$ are $N-1$ gapless transverse Goldstone modes, as depicted in Fig.~\ref{fig:MH}.
Equation (\ref{eq:action}) can be expanded into harmonic, anharmonic, and counter term parts, $S=\SH+\SAn+\SC$, 
\bea
\SH&=&{1\over 2g}\int_\sLambda \!\! d^{d+1} \!x~\Big[(\pz_\mu \sigma)^2+r^2 m_0^2\,\sigma^2 + (\pz_\mu \Bpi)^2 \Big], \label{actions}\\
\SAn&=&{m^2_0\over 2g}\int_\sLambda \!\! d^{d+1} \!x~\bigg[{ r\over \sqrt{N}}\,\big(\sigma^3+\sigma\Bpi^2\big)
+{1\over 4N}\,(\sigma^2+\Bpi^2)^2 \bigg], \ \nonumber \\
\SC&=&{(r^2-1) \,m_0^2\over 4g }\int_\sLambda \!\! d^{d+1} \!x~\Big[2r\sqrt{N}\, \sigma+\sigma^2+\Bpi^2\Big]\ ,  \nonumber
\eea
where the renormalized mass is $m =r m\nd_0$, and $r$ is determined by requiring $\langle \sigma \rangle = 0$ order by order in powers of $g$ and/or $1/N$. 
This counterterm prescription ensures that the Goldstone propagators remain massless at each order (see Appendix~\ref{app:counterterms}).

\subsection{Dynamical susceptibilities} 
Susceptibilities  are generally  defined by 
\beq
\chi\nd_{AB}(q) =\int \! d^{d+1}\!x~e^{iq\cdot x}\big\langle  A(x)\,B(0) \big\rangle\nd_{\rm c}\ ,
\eeq
where $A=\sigma$, $\Bpi$, $\Bpi^2$, {\it etc\/}. Here the subscript c denotes the connected average, $q=(\omega_n,\Bq)$ is the Euclidean momentum,
and $\omega_n$ is a bosonic Matsubara frequency.  The spectral function is $\chi_{AB}''(\Bq,\omega) = \textsf{Im} \,\chi\nd_{AB}(\Bq,\omega+ i0^+).$

In the broken symmetry phase, the longitudinal susceptibility as defined by (\ref{expand}) is
$\chi\nd_{\sigma\sigma}$. In contrast,  scalar fluctuations are given by
\beq
\rho(x)\equiv  {1\over \sqrt{N}}\left(|\BPhi(x)|^2 -  r^2 N\right)= 2 r\sigma +{\sigma^2+\Bpi^2\over \sqrt{N}}\ .
\eeq
One then defines the scalar susceptibility $\chi\nd_{\rho\rho}$, which is related to a sum of cross-susceptibilities,
\beq
\label{chiRR}
\begin{split}
\chi\nd_{\rho\rho}&= r^2 \big(4\chi\nd_{\sigma\sigma}+\chi\nd_{\rm sing} + \chi\nd_{\rm reg}\big), \\
\chi\nd_{\rm sing}  &=\frac{4}{ r\sqrt{N}} \, \chi\nd_{\sigma\Bpi^2}+\frac{1}{ r^2 N } \, \chi\nd_{\Bpi^2 \Bpi^2},\\
\chi\nd_{\rm reg}&=\frac{1}{ r^2 N } \, \chi\nd_{\sigma^2\sigma^2}+\frac{4}{ r\sqrt{N}} \, \chi\nd_{\sigma^2\sigma}+
\frac{2}{ r^2 N } \, \chi\nd_{\sigma^2\Bpi^2}.
\end{split}
\eeq
Anharmonic interactions  $\SAn$ in Eq.~(\ref{actions}) are responsible for  cross-susceptibilities between the
$\sigma$ and $\Bpi$ fields.   In Secs.~\ref{sec:weakCoupling} and \ref{sec:largeN} we  calculate  the longitudinal and scalar susceptibilities
using weak coupling and  large $N$ expansions, respectively.  Both approaches show that in the $\omega\to 0$ (infrared) limit,  
$\chi_{\sigma\sigma}''$ and $\chi_{\rm sing}''$ are singular,  while $\chi_{\rm reg}''$ is infrared regular.

\subsection{Conductivity}
For all  $N\ge 2$,  currents and conductivities can be defined as derivatives with respect to matrix gauge fields introduced into  Eq.~(\ref{eq:action}) by setting $\pz\nd_\mu\BPhi\to\pz\nd_\mu\BPhi + e A\nd_{a\mu} T^a \BPhi$.  The $T^a$ are the $O(N)$ symmetry generators and $e$ is the charge.\cite{QPT}

When the $O(N)$ symmetry is broken, the generators $T^a$ fall into two classes: broken and unbroken.  The $N-1$ broken generators rotate
between $\sPhi\nd_1\equiv\sigma$ and $\sPhi\nd_j\equiv\pi\nd_{j-1}$, with $j=2,\ldots,N$.  The remaining $\half (N-1)(N-2)$ unbroken generators rotate
among the $(N-1)$ components of the $\Bpi$ field.  In analogy with the $O(2)$ case, which has a single generator of the first kind 
($T = {i\over 2}\sigma^y$), we define the (paramagnetic) currents of the broken generators by
\beq
I^\textsf{para}_{\Sb\mu}(x)={e\over g}\,\pz\nd_\mu\BPhi\cdot T^\Sb\BPhi \ ,
\label{current}
\eeq
their correlators as
\beq 
K^\textsf{bb}_{\mu\nu}(x,x') = \big\langle I^\textsf{para}_{\Sb\mu}(x)\,I^\textsf{para}_{\Sb\nu}(x')\big\rangle\ ,
\eeq 
and the generalized $O(N)$ conductivity as
\beq 
\sigma(\omega)=\textsf{Im} \,{1\over\omega+i\varepsilon}\left( K^\textsf{bb}_{xx}(\omega+i0^+,\Bq=0) -
{e^2\over g}\big\langle |\BPhi|^2  \big\rangle\right) \ ,
\label{sigma}
\eeq
where the last term is the diamagnetic contribution.  The $O(N)$ conductivity is discussed more thoroughly in Appendix ~\ref{app:ONconductivity}.

\section{Weak coupling limit  ($g\ll 1$)}
\label{sec:weakCoupling}

\begin{figure}[!t]
\begin{center}
\includegraphics[width=6cm,angle=0]{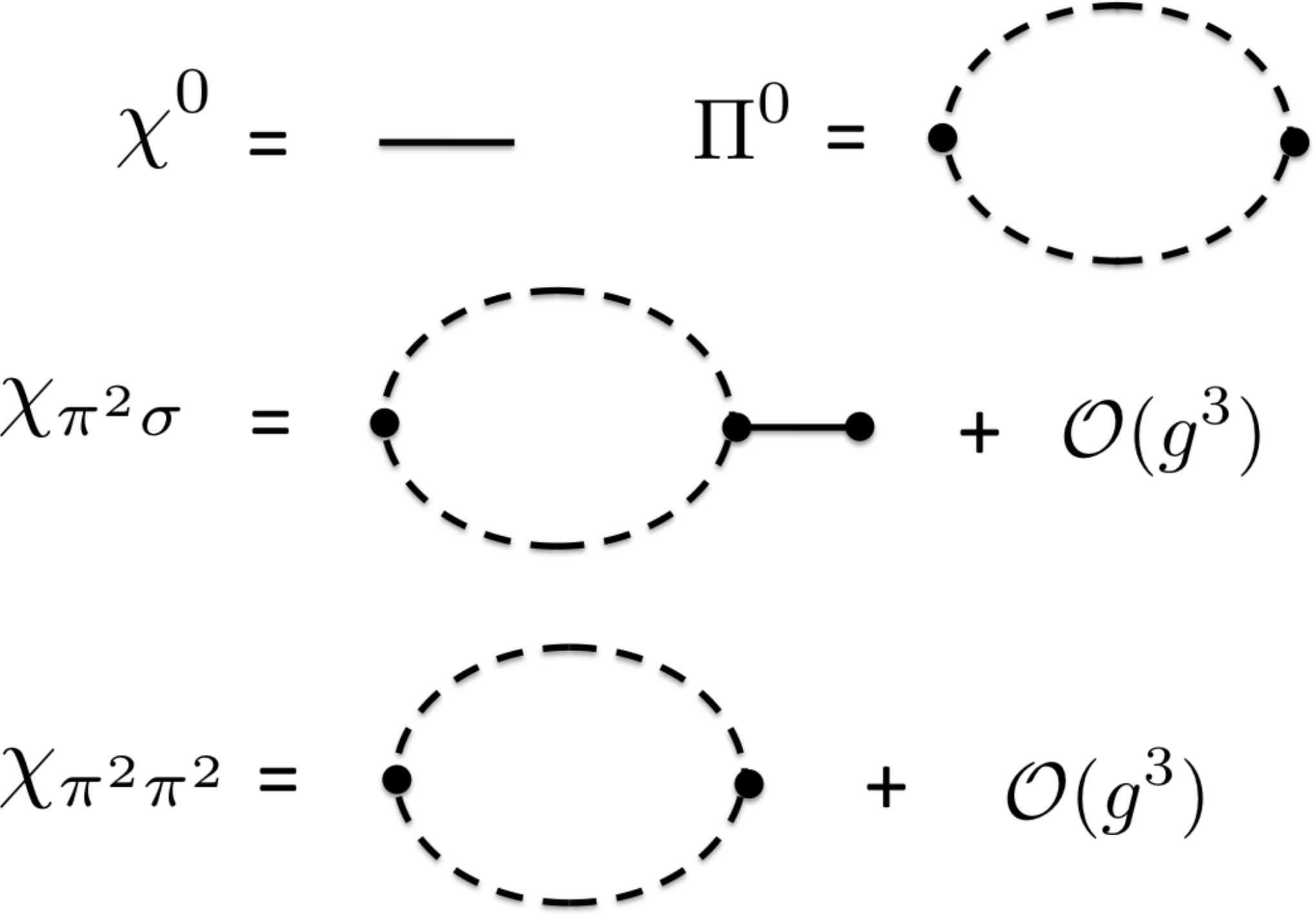}
\caption{{\bf Infrared divergences at weak coupling}. Diagrams describing the infrared divergent weak coupling corrections to the longitudinal and scalar susceptibilities  in Eqs.~(\ref{chiRR}) and (\ref{chiss}).  
Solid lines are $\sigma$ propagators; dashed lines represent $\Bpi$ propagators.\label{fig:weak}}
\end{center}

\end{figure}

Diagrammatic perturbation theory for $\chi$ in powers of $g$ amounts to counting loops.  To one loop order, 
\beq
\chi\nd_{\sigma\sigma}(q)=\chi^0(q)+\chi^0(q)\sPi^0(q)\chi^0(q) +{\cal O}(g^3)
\label{chiss}
\eeq
where $\chi^0=g/(q^2+m^2)$ is the zeroth order longitudinal susceptibility  and   $\sPi^0$ is the polarization bubble, shown in Fig.~\ref{fig:weak}. Since  $\sPi^0$
is a convolution of  two massless $\Bpi$ propagators, it diverges as (Appendix \ref{app:weakCoupling})
\beq
\label{Pi0}
\begin{split}
\sPi^0(q)&= \frac{m_0^4r^2(N-1)}{2N}\!\!\int\!\!{d^{d+1}\!k\over(2\pi)^D} {1\over k^2(k+q)^2}  \\
&=\frac{m_0^4r^2(N-1)}{N} \begin{cases}
{1\over 16 \, |q|} & (d=2)\\
{1\over 32\pi^2} \left[ 1+\ln\left({\sLambda^2\over q^2}\right)\right] &  (d=3) \ .
\end{cases}
\end{split}
\eeq 
We ignore all other diagrams of the same order  in $g$ which do not contribute to the low-frequency dependence, such as
the loop  of massive $\sigma$ propagators.  Analytically continuing  $|q|\to \sqrt{\Bq^2 -(\omega+i\epsilon)^2}$ and taking the imaginary part yields
\bea
\chi''_{\sigma\sigma}&=&\ {\pi g\over 2\sqrt{\Bq^2+m^2}}\,\delta\big(\omega-\sqrt{\Bq^2+m^2}\big)  \nonumber\\
&&\qquad + g^2\>{(N-1) \, m_0^4r^2\over 32\pi N} \>  {\Theta(\omega^2-\Bq^2)\over (\Bq^2+m^2-\omega^2)^2}\nonumber\\
&&\qquad\qquad\quad \cdot \begin{cases} {2\pi\over\sqrt{\omega^2-\Bq^2}} & (d=2) \\ \quad\  1& (d=3)\ .\end{cases} \label{chiSSexp}
\eea
Note that $\chi''_{\sigma\sigma}$ at zero momentum behaves as $\omega^{d-3}$, which is a direct consequence of the low-momentum
divergence of $\sPi^0(q)$.  This divergence is the quantum version of the divergent longitudinal susceptibility of the $O(N)$ ferromagnet
($N>2$) in $d+1$ dimensions in its low-temperature ordered phase.\cite{PP}
 
The leading order  corrections  of $\chi\nd_{\rho\rho}$  are given by the terms in Eq.~(\ref{chiRR}) which are  dominated by infrared-divergent $\sPi^0$ factors.
The dominant terms are depicted in Fig.~\ref{fig:weak}.
Computing the scalar susceptibility up to one loop yields
\beq
\chi\nd_{\rho\rho}= {4g r^2\over q^2+m^2}+ {4 g^2 q^4 r^2\over \left(q^2+m^2\right)^2}\>\sPi^0(q)+r^2\chi\nd_{\rm reg}(q) \ .
\label{chiRRexp}
\eeq 
Note that the infrared singularities  cancel out,  leaving  $\chi\nd_{\rho\rho}$ to rise as  $q^4 \sPi^0(q)$ at low $q$.
As a consequence the scalar susceptibility decays rapidly at low frequencies,
\beq
\chi_{\rho\rho}''\sim \begin{cases}
(\omega^2-\Bq^2)^{3/2}\,\Theta(\omega-|\Bq|) & (d=2) \\
(\omega^2-\Bq^2)^2\ln\!\big |\omega-|\Bq|\big| & (d=3)\ ,
\end{cases}
\label{chirr-g}
\eeq
which enables a pronounced amplitude mode peak at $\omega\sim~m$ due to the mass pole of $\chi^0$.
Note that when the action is parametrized in terms of $\rho$ and $\nhat$, the amplitude-direction coupling $\rho(\partial_\mu\nhat)^2$ gets two extra derivatives relative to the longitudinal-transverse coupling $\sigma\Bpi^2$.  This is responsible for infrared suppression by a factor $\omega^4$ between Eqns. (\ref{chiSSexp}) and (\ref{chirr-g}). This behavior 
is also seen at large $N$, as derived in Sec.~\ref{sec:largeN} and shown in Fig.~\ref{chifig}.

\begin{figure}[!t] 
\centering
\includegraphics[width=7.0cm,angle=0]{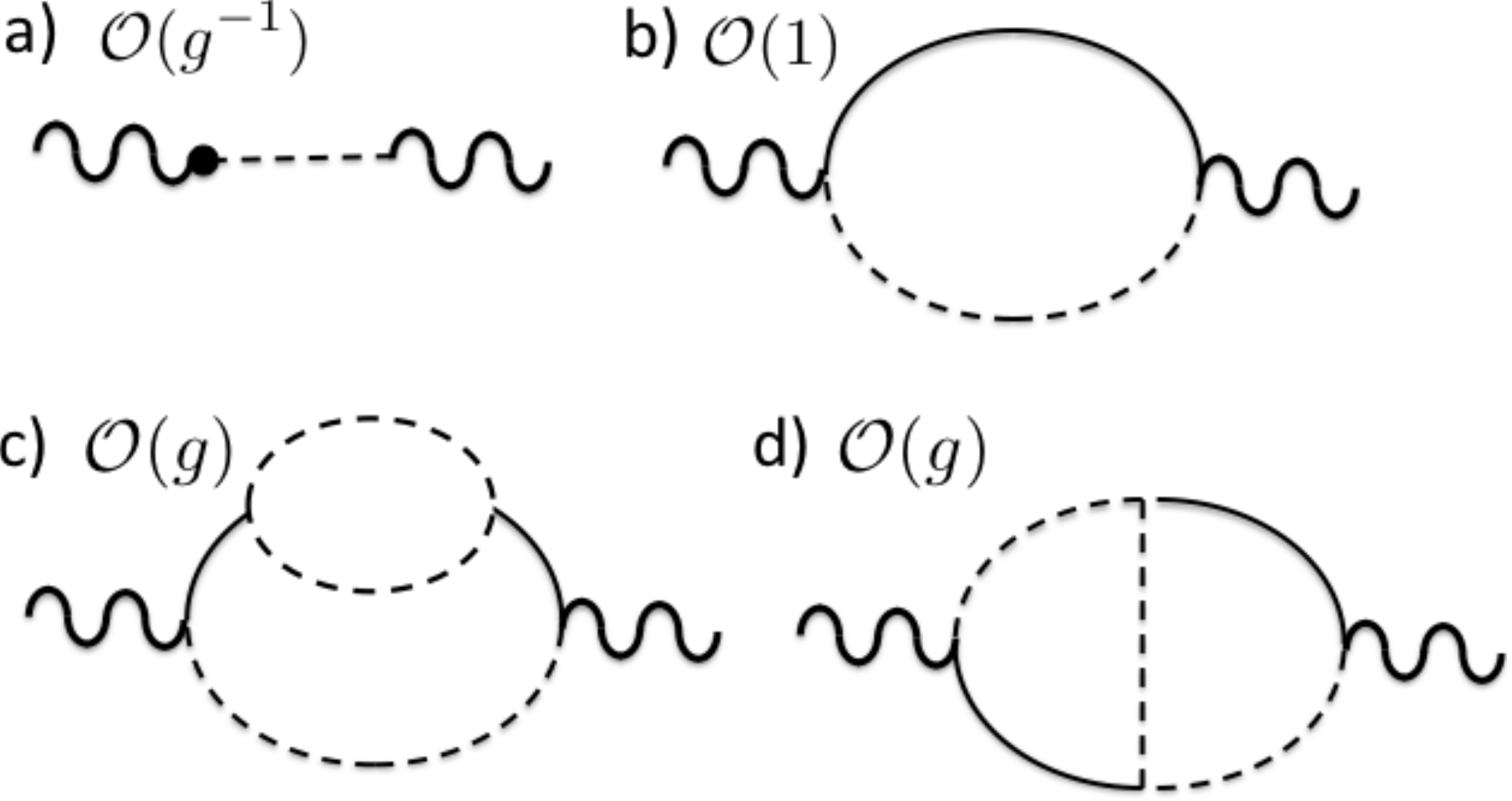}
\caption{ 
{\bf Weak coupling expansion for the dynamical conductivity.}  (a) Order $g^{-1}$ contribution to the weight of the superfluid delta function at $\omega=0$.
(b) Order $g^0$ diagram has a threshold at the amplitude mode mass. (c),(d) Two-loop self-energy and vertex  corrections which contribute to the subgap conductivity.
As $N\to\infty$ diagram (c) dominates and yields $\sigma\sim \omega^{2d-3}$.  
For $N=2$ (the relativistic superfluid), cancellations between diagrams (c) and (d)  suppress the subgap conductivity by four powers of $\omega$,
and $\sigma\sim \omega^{2d+1}$ (see text).} \label{fig:sigma-diagrams} 
\end{figure}

\subsection{Conductivity} 
\begin{figure}[!t] 
\begin{center}
\includegraphics[width=7.5cm,angle=0]{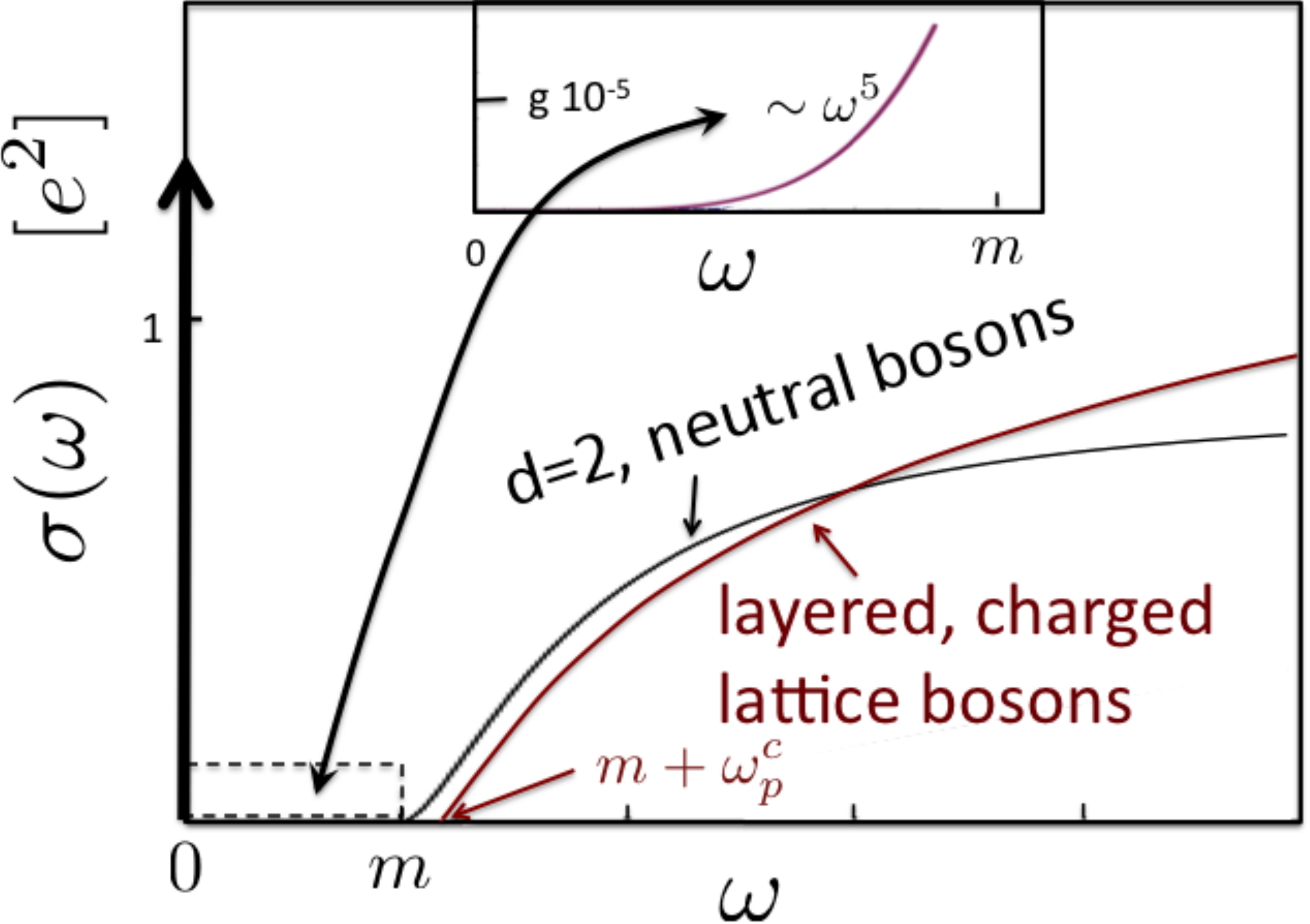}
\caption{{\bf Dynamical conductivity for $O(2)$ (relativistic bosons) in two dimensions}.  The arrow at zero frequency denotes is the superfluid
delta function response.  For neutral bosons, there is a weak  ${\cal O}(g)$ sub-gap tail (see inset).  The conductivity of a
bosonic layered superconductor (red online) is plotted for using intralayer plasma frequency  $\omega_{\rm p}=10\,m$ and a much smaller
interlayer plasma  frequency of  $\omega^c_{\rm p}=0.1\, m$. }
\label{sigmafig} 
\end{center}
\end{figure}

Diagrams contributing to $\sigma(\omega)$ are depicted in Fig.~\ref{fig:sigma-diagrams}.
The leading order conductivity (\ref{sigma})   is of order $g^{-1}$ and is all contained in the delta function weight at zero frequency,
\beq
\sigma(\omega) = A\,\delta(\omega) + \stil(\omega)\,
\eeq
where $A=N e^2 g^{-1}+\CO(g^{0})$.
A nontrivial frequency dependence arises at
order $\CO(1)$ from the $\sigma$-$\Bpi$ bubble diagram shown in Fig.~\ref{fig:sigma-diagrams}. It exhibits a power law rise above a mass gap threshold:
\beq
\stil\nd_0(\omega)={\pi S\nd_d e^2\over d\omega^2} \bigg( {\omega^2-m^2\over 4\pi\omega}\bigg)^{\!d} \>\Theta(\omega^2-m^2)\ ,
\eeq
where $S\nd_d$ is the surface area of a unit sphere in $d$ dimensions. The $\CO(1)$ threshold conductivity  is depicted in the main part
of Fig.~\ref{sigmafig}.    Other  $\CO(1)$ diagrams introduce frequency-dependent corrections at twice the mass gap, and overall renormalizations
of the superfluid density $A$ and the mass gap $m$.

Two-loop diagrams are of order $\CO(g)$. The two diagrams which produce finite sub-gap conductivity  are depicted in the bottom of 
Fig.~\ref{fig:sigma-diagrams}.  As $N\to\infty$, diagram (c) dominates the  sub-gap conductivity.  By power counting, the diagram scales as
$\sigma \sim \omega^{2d-3}$, implying a significant sub-gap absorption for $d=2$ for the large $N$ conductivity.  

At smaller $N$, however, diagram (d) becomes comparable to (c) but opposite in sign, tending to cancel the sub-gap conductivity. After a lengthy calculation, shown in Appendix \ref{app:condCalculation}, we obtain the power series in frequency, 
\beq
\label{subgap}
\begin{split}
\stil^{d=2}_g&=\frac{g e^2 m}{2^{8}N\pi}\left\{\scriptstyle{(N-2)}\left(\frac{16\omega}{15m}+\frac{32\omega^3}{105 m^3}\right) +
\scriptstyle{(3N-5)}\frac{16 \omega^5}{315 m^5}  + \ldots \right\},  \\
\stil^{d=3}_g&=\frac{g e^2 m^3}{3\pi^2 2^{9}N}\left\{ \scriptstyle{(N-2)}\left(\frac{\omega^3}{4m^3}+\frac{\omega^5}{10m^5}\right) +
\scriptstyle{(9N-16)}\frac{\omega^7}{180 m^7} +\ldots\right\}.
\end{split}
\eeq

Remarkably,  we find that {\em for $N=2$, the coefficients of the two lowest powers vanish! } This result can be understood as a consequence of the complete $O(2)$ symmetry
of the conductivity, which implies that it does not excite Goldstone fluctuations, similar to the scalar susceptibility. In contrast for $N>2$, the conductivity is not a pure scalar 
response,  since it  depends explicitly on the broken symmetry direction.

\section{Large $N$ limit}
\label{sec:largeN}

The inverse number of components $1/N$ controls an expansion about  the  $N=\infty$ limit for all values of $g$.
This allows us to approximate the finite $N$ system in both
ordered and disordered phases, except close to the quantum phase transition at $g_{\rm c}$. 
Note that all anharmonic terms in $\SAn$ are suppressed by negative powers of $N$.

The large $N$ renormalization of the order parameter is (see Appendix \ref{app:counterterms})
\beq
r^2(g,\Lambda) =1- g\!\!\int \!\!{d^{d+1} q\over (2\pi)^{d+1} }{1\over q^2}\equiv 1-{g\over g^\infty_{\rm c}}\ ,
\label{mRenorm}
\eeq
with $r^2$ vanishing linearly with $g$ at the quantum critical point,
\beq
g^\infty_{\rm c}=\begin{cases}
4\pi/\sLambda & d=2 \\
8\pi^2\!/\!\sLambda^2 & d=3\ ,
\end{cases}
\label{gc}
\eeq
The renormalized mass $m^2=m_0^2\,(1-g/g^\infty_{\rm c})$ also vanishes at $g^\infty_{\rm c}$.

Now we evaluate the large $N$ longitudinal susceptibility, given by
\beq
\chi^\infty_{\sigma\sigma} (q)={g\over q^2+m^2-g\sSigma_\sigma(q) } \ ,
\eeq
where
$\sSigma_\sigma$ is the longitudinal self-energy given by the RPA sum~\cite{Subir},
\beq
\sSigma_\sigma(q)={\sPi^0(q)\over 1 + g \, \sPi^0(q)/m^2},
\eeq
as shown in Appendix \ref{app:RPA}.
Note that, since $\sPi^0(q)$ diverges as $q\to 0$, $\Sigma_\sigma(0)=m^2/g,$ in agreement with an exact Ward identity.\cite{Nepomnyashchii,Dupuis} As a consequence, the pole in $\chi_{\sigma\sigma}$ at $q^2=m^2$  gets replaced by a branch cut starting at $q^2=0$.  For instance, for $d=2$ we obtain
\beq
\chi_{\sigma\sigma}(q)={g\over q^2+{16\,|q| m^2 \over g m_0^2 + 16\, |q|}}\ . 
\label{chiss2d} 
\eeq
This infrared singularity, given at large $N$ to all orders in $g$, agrees with the singularity obtained for all $N$  at   order $g^2$ in  Eq.~(\ref{chiSSexp}).

Now we evaluate the scalar susceptibility [Eq.~(\ref{chiRR})],
\beq
\chi_{\rho\rho}=4r^2\chi\nd_{\sigma\sigma}+ r^2\chi\nd_{\rm sing} + \CO(1/N).
\eeq
$\chi\nd_{\sigma\Bpi^2}$ and $\chi\nd_{\Bpi^2\Bpi^2}$ are given by
\beq
\frac{4}{ r\sqrt{N}} \> \chi_{\sigma\Bpi^2}(q)=-{4g^2\sSigma_\sigma(q)/m^2 \over q^2+m^2-g\sSigma_\sigma(q)}
\eeq
and
\beq
\frac{1}{r^2 N}\>\chi_{\Bpi^2\Bpi^2}(q)={4g^2(q^2+m^2)\sSigma_\sigma(q)/m^4 \over q^2+m^2-g\sSigma_\sigma(q)}\  .
\eeq
Summing all the contributions in Eq. (\ref{chiRR}) yields
\bea
\chi\nd_{\rho\rho}(q)={4gr^2 \over q^2+m^2}\left(1+{g \, q^4\sSigma_\sigma(q)/m^4 \over q^2+m^2-g\sSigma_\sigma(q)}\right)  \ .  \label{chiRRN}
\eea
Note that the factor of $q^4$ in the numerator suppresses the low $q$ singularity in the denominator, and  $\chi_{\rho\rho}(q)\propto |q|^{d+1}$
at  low momenta, just as in the weak coupling case. In Fig.~\ref{chifig} the large $N$ approximations for
 $\chi_{\sigma\sigma}''(\omega)$ and $\chi_{\rho\rho}''(\omega)$ are plotted for $d=2$ . We take $g/g_{\rm c}^\infty(\sLambda)=0.84$, inside the ordered phase.  The  amplitude mode peak is  clearly visible in $\chi_{\rho\rho}''$ while it is difficult to detect in  $\chi_{\sigma\sigma}''$.

\subsection{Width of the scalar peak}

Thus far we have compared the scalar and longitudinal susceptibilities and have shown that a peak can be discerned  in $\chi_{\rho\rho}''(\omega)$  even in cases where it is hidden by infrared divergences in $\chi_{\sigma\sigma}''(\omega)$ (see Fig.~\ref{chifig}).  Deep inside the ordered phase,  $g\ll g_c$, the peak in $\chi_{\rho\rho}''(\omega)$ becomes very sharp in relation to its energy $\omega=m$.  However, as one approaches the disordered phase, $g\to g_c$, the relative width grows until close enough to the transition the peak in $\chi_{\rho\rho}''(\omega)$ becomes broader than $m$.  Thus, close enough to the transition, it becomes impossible to identify the Higgs energy from  $\chi_{\rho\rho}''(\omega)$.    

We can study the width of the peak systematically in the large $N$ limit.  The scalar susceptibility at ${\bf q}=0$ can be written as
\begin{eqnarray}
\chi_{\rho\rho}''(\omega)={4g\over m_0^2}{\omega^4 {\rm Im} \left[F^*(\omega)\right]\over \left| \omega^2+(\omega^2-m^2)F(\omega) \right|^2},
\end{eqnarray}
where
\begin{eqnarray}
F(\omega)\equiv{m^2\over g\Pi^0(-i\omega)}.
\end{eqnarray}
For example, from Eq.~(\ref{Pi0}) we find that for $d=2$ and $N=\infty$, $F(\omega)=-i\omega/(2\gamma)$, where
\begin{eqnarray}
\gamma={g m_0^2\over 32}.
\end{eqnarray} 
Thus,
\begin{eqnarray}
\begin{split}
\chi_{\rho\rho}''(\omega)={4 g\over m_0^2}{2\gamma \omega^3 \over(\omega^2-m^2)^2+4\gamma^2\omega^2},
\end{split}
\label{chiRhoGamma}
\end{eqnarray}
For $\gamma<m$, Eq.~(\ref{chiRhoGamma}) is peaked at $\omega=m$ with width smaller than $m$.  On the other hand, for $\gamma>m$, the width becomes larger than $m$ and the peak is shifted to energies larger than $m$.   Hence, $\gamma=m$ marks the
point beyond which the Higgs mass cannot be determined accurately from the peak in $\chi_{\rho\rho}''(\omega)$.   

\begin{figure}[!t]
\centering
\includegraphics[width=8cm]{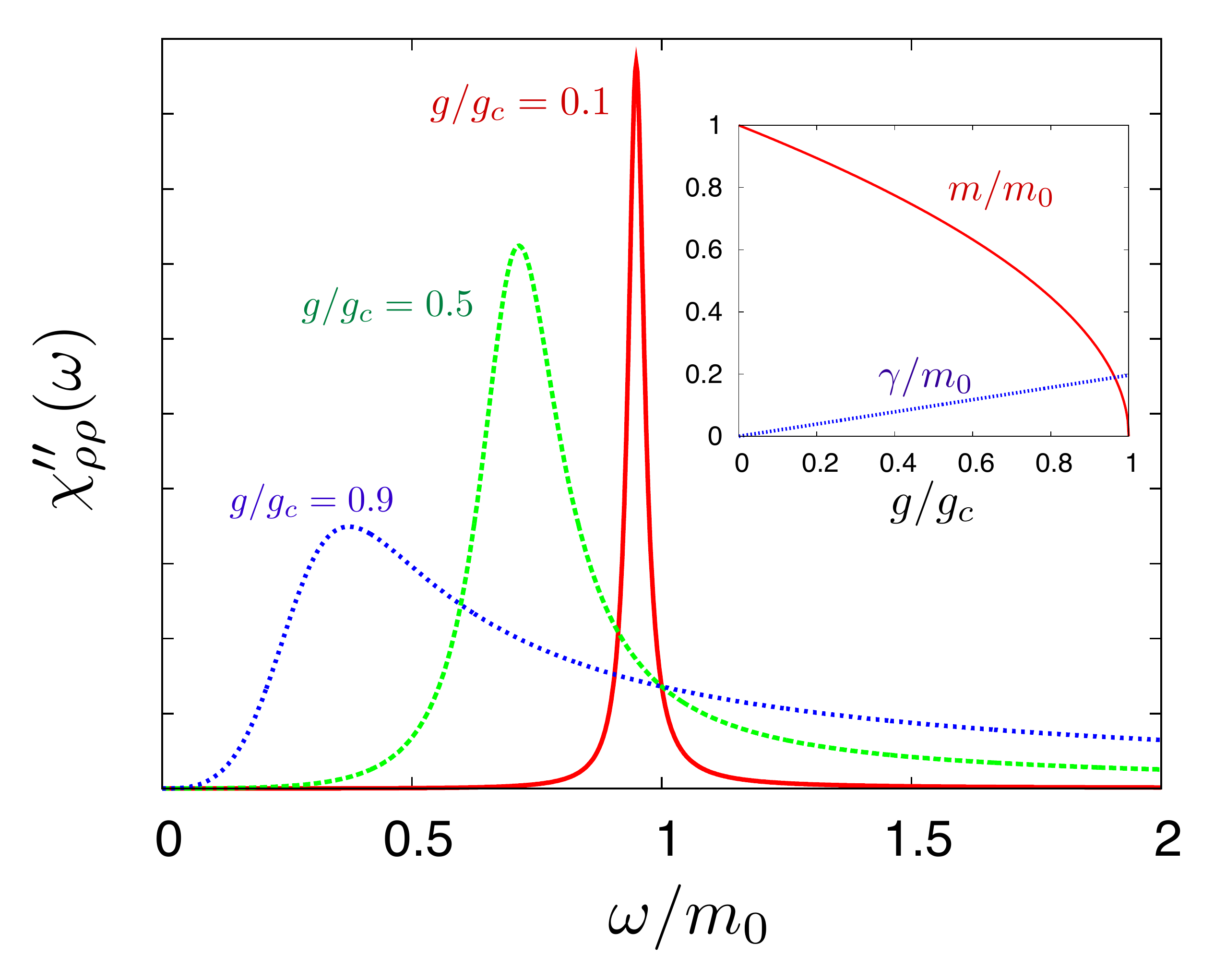}
\caption
{\label{fig:msharp} {\bf Dependence of the scalar peak on $g$}.  Scalar susceptibility $\chi_{\rho\rho}''(\omega)$ for different values of $g/g_c$ in $d=2$ dimensions.  Results shown in the large $N$ limit for $\Lambda=2 m_0$. (Inset) Renormalized mass $m$ (solid curve) and width $\gamma$ (dashed curve) of the peak in $\chi_{\rho\rho}''(\omega)$ as a function of the tuning parameter $g$, expressed in units of the bare mass $m_0$.  For $g\to 0$, deep inside the ordered phase, the mass approaches its bare value $m_0$ and the peak is very sharp, $\gamma\to 0$.  As $g$ approaches $g_c$, the mass is softened and the width grows, until for $g/g_c>0.96$ the peak energy is smaller than the width.}
\end{figure}

Note that $\gamma$ grows linearly with $g$, whereas $m=m_0 r$ vanishes at $g=g_c$, according to Eq.~(\ref{mRenorm}).  This is depicted in Fig.~\ref{fig:msharp}, where it is seen that close enough to $g_c$ the width of the peak $\gamma$ exceeds the renormalized Higgs mass $m$. The question of how close one can get to the transition before this happens is nonuniversal, since it depends on the ratio $\Lambda/m_0$.  For $\Lambda=2 m_0$, this occurs at $g/g_c=0.96$, corresponding to $m/m_0=0.19$.  For larger values of $\Lambda/m_0$, the softening of the Higgs mode can be tracked to very low energies before its mass can no longer be detected reliably. 
 
For $d=3$ and $N=\infty$, we obtain
\begin{eqnarray}
\chi_{\rho\rho}''(\omega)={4g\over m_0^2}{\eta^2 \pi \omega^4 {\rm sign}(\omega)\over \left[\omega^2-m^2+\eta^2 \omega^2\left(1+\log\frac{\Lambda^2}{\omega^2}\right)\right]^2+\eta^4\pi^2\omega^4 },\nonumber\\
\end{eqnarray}
where
\begin{eqnarray}
\eta^2={g m_0^2\over 32 \pi^2}.
\end{eqnarray}
In this case, in contrast to $d=2$, a sharp peak can be observed arbitrarily close to the critical point $g_c$, that is, for arbitrarily small $m/m_0$.

\section{Experimental Probes}
\label{sec:experiments}
\subsection{Lattice bosons near the Mott transition}
Ultracold bosons in an optical lattice undergo a Mott transition  at integer fillings as the lattice potential is strengthened.\cite{Fisher}   
The transition is well described by an $O(2)$ relativistic field theory,\cite{EAAA} with the lattice strength controlling the radius of the Mexican hat.
Amplitude oscillations could, in principle, be observed after quenching the Mott phase into the superfluid phase, similar to coherence peaks recovery seen
by Greiner {\em et al.},\cite{Greiner} who studied the opposite quench direction.
More directly, by modulating the lattice potential at some frequency $\omega$, one can excite the scalar mode of the superfluid.\cite{Huber} Thus, the system is predicted to absorb energy at a rate $\omega\,\chi''_{\rho\rho}(\omega).$  Such measurements have indeed been carried out,\cite{Esslinger} and although no sharp peaks were observed, this can likely be attributed to the presence of a harmonic trapping potential, whose effect is to smear the energy of the amplitude mode, and the application of long-lasting modulation pulses that pushed the system outside of the linear response regime.

The phase of the optical lattice can also be modulated instead of its amplitude.  The energy absorption rate is then given the optical conductivity $\sigma(\omega)$ at the phase modulation frequency \cite{Giamarchi} which, as shown above, has a threshold at the Higgs mass.

Finally, recent experiments using Bragg spectroscopy \cite{Sengstock} have shown evidence of the amplitude mode in interacting lattice bosons.  Although these experiments were carried out beyond the linear response regime treated here, they demonstrate that the Higgs mode is, in principle, observable though the use of these probes.

\subsection{Raman scattering in Antiferromagnets and Charge Density Wave Systems}
Heisenberg antiferromagnets, and incommensurate charge density wave systems can be effectively described by relativistic $O(3)$ and $O(2)$ theories,
respectively.  Inelastic light scattering \cite{Raman,Review} can effectively couple to the square of the local order parameter.
For example, in antiferromagnetic insulators, light couples to bond spin operators,\cite{Fleury,2-Mag}
\beq
\label{HRaman}
\begin{split}
H\nd_{\rm Raman} &=D\sum_{\Bx,\Beta}   E\nd_\Beta \, E'_\Beta \, \BS\nd_\Bx \cdot \BS\nd_{\Bx+\Beta}\\
& \propto \sum_\Beta\int \!\! d^d \!x \> E\nd_\Beta \, E'_\Beta \,  \Big( |\BPhi(\Bx)|^2+\ldots  \Big)\ .
\end{split}
\eeq
Thus, with incoming (outgoing) electric field polarization $\BE$ ($\BE'$), consistent with  the crystal symmetry, the Raman spectrum 
measures the scalar susceptibility $\chi''_{\rho\rho}(\omega)$.  The small but finite separation between the two spin operators leads to small corrections, denoted by $\ldots$ in Eq. (\ref{HRaman}).   The most important of these corrections, proportional to $(\Beta\cdot\nabla \Bpi)^2$ gives a direct coupling of light to pairs of spin waves and leads to a broad background signal.  The background is well-behaved in the infrared due to the extra spatial derivatives in the coupling and hence is not expected to hide any peaks present in $\chi''_{\rho\rho}(\omega)$.

Indeed, a pronounced Raman peak has been measured in several magnetic compounds.\cite{Raman-exp}
Theoretically, it has been analyzed as a two-magnon resonance, or bound state,\cite{2-Mag} which is equivalent (in its quantum numbers) to the
amplitude mode.  We note that inelastic neutron scattering probes the longitudinal susceptibility of antiferromagnets, since the neutron spin couples locally to the N\'eel vector. At the Bragg  wave vectors,  the  amplitude mode peak  is therefore expected to be obscured by the singular low-energy scattering.

For incommensurate charge density wave the order parameter has a massive amplitude mode, and a soft translational mode. Light excites the amplitude mode by inducing inter-band transitions.\cite{CDW-Raman-theory}  This has been used to detect the amplitude mode in low-dimensional CDW systems through Raman scattering \cite{CDW-Raman-exp} and femtosecond pump-probe spectroscopy \cite{CDW-PP,CDW-PP2,CDW-PP3} experiments.  In contrast to antiferromagnets, in the case of CDW order, neutrons couple to the local charge density and hence act as a scalar measurement.   Indeed, the amplitude mode of a CDW has been measured using neutrons.\cite{CDW-neutrons}  Similarly, neutrons have been proposed to detect the amplitude mode of a DDW state.\cite{DDW}

\subsection{Superconductors}
Granular superconducting films and  low-capacitance Josephson junction arrays which exhibit superconductor to insulator transitions, can be effectively described in terms of a   bosonic  $O(2)$ relativistic field theory.\cite{Fisher} 
By proximity to a Mott insulating phase, the  amplitude mode may be suppressed below  the BCS pairing gap, and appear as a long-lived collective excitation.

In homogenous BCS superconductors, the detection of the amplitude mode by Raman scattering has been proposed in the presence of a coexistent charge density wave.\cite{LittlewoodVarma} Here we propose that the amplitude mode may be observed in a class of ``bosonic'' superconductors,
that is, those exhibiting   short coherence length, low superfluid density, and perhaps a pairing gap above $T_{\rm c}$ ({\it e.g.}, in cuprates~\cite{Uemura,EK}).
Such superconductors may be  described by charged lattice bosons, which may be treated by the $O(2)$ theory of Eq.~(\ref{eq:action}) with
long-range Coulomb interactions.  The optical conductivity of HCB couples directly to the amplitude mode;\cite{LA-PRB} however, since
Cooper pairs are charged, long ranged Coulomb interactions modify  the Goldstone mode's dispersion.  In a three-dimensional sample,
the phase fluctuations are gapped at the plasma frequency  $\omega_{\rm p}$.

Nevertheless,  in highly anisotropic  layered superconductors such as the cuprates,\cite{MA} the threshold for optical absorption at zero temperature 
is shifted from $m$ by the  relatively small $c$-axis plasma frequency $\omega^c_{\rm p}\ll \omega_{\rm p}$, as shown in Appendix \ref{app:LLB}.   We propose that
Raman scattering, which couples to the $O(2)$  scalar susceptibility,  may be used to  observe the amplitude mode peak.
The mass $m$ of this mode is expected to be
of the order of the superfluid density and $T_{\rm c}$ and decrease toward the quantum phase transition into the insulating phase.

For example, in Bi$_2$Sr$_2$CaCu$_2$O$_{8+\delta}$, $\omega_{ab} \sim 1$\,eV, while
$\omega_{c} \sim 1$\,meV, leading to conductivity qualitatively similar to the charged latice bosons shown in Fig.~\ref{sigmafig}. 
We propose that the amplitude mode may be partially responsible for
the rise in optical conductivity in the mid infrared regime, as observed above 400 cm$^{-1}$ \cite{BasovTimusk}.

\section{Summary} 
We have calculated the scalar susceptibility and the conductivity by weak coupling large $N$ expansions within the SSB phase of the two- and three-dimensional $O(N)$ field theory.  In contrast to the longitudinal susceptibility, where low-frequency dissipation by Goldstone modes can entirely  mask the amplitude mode, the $O(N)$ symmetric susceptibility exposes the amplitude mode as a finite-width peak that is uncontaminated by infrared contributions arising from the Goldstone modes.   Similarly, the $O(2)$ conductivity exposes the amplitude mode as a broadened threshold in frequency.

From an operational point of view, the mass $m$ can be extracted experimentally from the peak in the $O(N)$ susceptibility.  While the determination of the peak energy is limited by the width of the peak, one can identify such a peak unambiguously as arising from a Higgs mode by tracking the peak position $m$ as a function of a tuning parameter near a quantum phase transition.  Deep in the ordered phase, the peak is very sharp relative to its energy $m$.  As one approaches the phase transition, the peak energy softens and its width grows.  Tracking the softening of the peak position is then possible except for a region very close to the transition, when the peak width becomes comparable to its energy.  Similar considerations apply to the broadened threshold in the $O(2)$ conductivity.

The suppression of sub-gap absorption by four powers of frequency was derived by precise cancellations between self-energy and vertex corrections
in the conductivity. It is easier to understand in the amplitude-direction representation, where 
$\rho$ is coupled to the derivatives of the order parameter direction $\rho\,(\pz\nd_\mu\nhat)^2$. Thus, the intrinsic dissipation of the $\rho$ self-energy
is suppressed by four powers of momenta. The current operators in all non-Abelian theories $N>2$ are not $O(N)$ symmetric, since they
contain  explicit coupling to the angle variables. For $O(2)$, however, the   current operator   $j_x=e \rho\,\partial_x\varphi$ is rotationally invariant and therefore
has four higher powers of sub-gap absorption than the non-Abelian conductivities. 

Our conclusion is therefore that  the amplitude mode is
in fact long-lived even for moderate $g<g_{\rm c}$, but its detection requires using $O(N)$ symmetric  experimental probes.
We propose Raman scattering for antiferromagnets and superconductors;
coherence peak oscillations and lattice modulation experiments in superfluids near the Mott transition; and optical conductivity in bosonic superconductors.

{\em Acknowledgements.}  We thank Ehud Altman,  Nicolas Dupuis, Rudi Hackl, Sebastian Huber, Steve Kivelson, Aneesh Manohar, 
Subir Sachdev, and Wilhelm Zwerger for useful discussions.  We acknowledge support from the Israeli Science Foundation, the U.S.-Israel Binational Science
Foundation, and the European Union grant IRG-276923.  One of the authors (DPA) was also supported by NSF grant DMR-1007028.
We are grateful to  KITP at Santa Barbara and Aspen Center For Physics where some of this work was initiated.

\appendix
\section{Counterterms in the broken symmetry phase }
\label{app:counterterms}
At small $g<g_c$, the order parameter acquires a finite vacuum expectation value (VEV) $\langle \BPhi_\sigma\rangle =  r \sqrt{N} $ in the $\sigma$ direction. We choose to expand the Euclidean-time action about the VEV 
\beq
\BPhi=( r\sqrt{N}+\sigma \, , \, \Bpi)\ ,
\label{ap:expand}
\eeq
which leads to the action
\bea
S&=&\SH+\SAn+\SC\nonumber\\
\SH&=&{1\over 2g}\int_\sLambda \!\! d^{d+1} \!x~\Big[(\pz_\mu \sigma)^2+r^2 m_0^2\,\sigma^2 + (\pz_\mu \Bpi)^2 \Big], \label{action1}\\
\SAn&=&{m^2_0\over 2g}\int_\sLambda \!\! d^{d+1} \!x~\bigg[{ r\over \sqrt{N}}\,\big(\sigma^3+\sigma\Bpi^2\big)
+{1\over 4N}\,(\sigma^2+\Bpi^2)^2 \bigg], \ \nonumber \\
\SC&=&{(r^2-1) \,m_0^2\over 4g }\int_\sLambda \!\! d^{d+1} \!x~\Big[2r\sqrt{N}\, \sigma+\sigma^2+\Bpi^2\Big]. \nonumber  
\label{ap:actions}
\eea
The harmonic action $S\nd_0$ provides zeroth-order massive and massless propagators,
\bea
G^0_{\sigma\sigma}&=& {g \over k^2 + m^2 }\nonumber,\\
G^0_{\Bpi_i \Bpi_j}&=&{g \over k^2} \>  \delta_{ij},
\eea
where $m=r\,m_0$ is the renormalized mass. 

The parameter $r$ is chosen such that that the VEV of $\sigma$ is zero, that is to say, that the fields $\sigma$ and $\Bpi$ are expanded about one of the true ground states of the system.  This is equivalent to requiring the vanishing of the $\sigma$ tadpole; that is, the sum of all 1PI diagrams with a single external $\sigma$ line must vanish. 

In the large $N$ limit, we can compute $r$ in closed form.  At leading order in $N$, ${\cal O}(\sqrt{N})$, there are only two 1PI diagrams: a sigma line terminating in a $\Bpi$ loop,
\beq
-{m^2_0\, r\,(N-1)  \over 2g\sqrt{N}}\int^\sLambda{d^{d+1} \Bk\over (2\pi)^{d+1}} {g\over k^2},
\eeq
and a $\sigma$ line ending in a counter term vertex,
\beq
-{m_0^2\,r\,(r^2-1) \sqrt{N}\over 2g }.
\eeq
Setting the sum of the two terms to zero yields (for $N\to\infty$),
\bea
r^2&=&1-g\int{d^{d+1} \Bk\over (2\pi)^{d+1}} {1\over k^2}  +\CO \left({1\over N}\right)\nonumber\\
&=&1-g/g_c^\infty.
\eea
Evaluating the integral with a cutoff $\Lambda$ on the spatial momenta (but no cutoff on the Matsubara frequencies) yields 
\beq
g^\infty_{\rm c}=\begin{cases}
4\pi/\sLambda & d=2, \\
8\pi^2\!/\!\sLambda^2 & d=3.
\end{cases}
\eeq

Since we expand the fluctuations about the true ground state, Goldstone's theorem guarantees that $\Bpi$ field is massless. Indeed, in the large $N$ limit, the $\Bpi$ self-energy is the sum of two diagrams, a $\Bpi$ loop and a counterterm:
\bea
\Sigma_{\Bpi_i\Bpi_j}&=&-\delta_{ij} {m_0^2\over 2g}\left[\int^\sLambda{d^{d+1} \Bk\over (2\pi)^{d+1}} {g\over k^2}+(r^2-1)\right]+\CO \left({1\over N}\right)\nonumber\\
&=&0.
\eea
Thus, for $N=\infty$, the bare and renormalized $\Bpi$ propagators are identical to each other.

A similar computation shows that the constant (momentum-independent) contribution to the $\sigma$ self-energy cancels for $N=\infty$, thus identifying $m=m_0 r$ as the renormalized mass of the amplitude mode beyond tree level.

In the weak coupling regime, all of our computations are carried out to leading nontrivial order in $g$.  In this case, we can set $r=1$ and also ignore the counterterms, which only correct our results at subleading orders in $g$.

\section{Computing the longitudinal susceptibility in the weak coupling limit $g\ll 1$}
\label{app:weakCoupling}

In momentum space, the bare susceptibility $\chi^0_{\sigma\sigma}(q)$ is
\begin{equation}
\chi^0_{\sigma\sigma}(q)={g\over q^2+m^2}\ ,
\end{equation}
where $q^2\equiv q\nd_\mu\,q\nd_\mu$.  The full susceptibility is
\begin{equation}
\chi\nd_{\sigma\sigma}(q)={1\over \left[\chi^0_{\sigma\sigma}(q)\right]^{-1}-\Sigma\nd_\sigma(q)}\ .
\end{equation}
Expanding, we have
\begin{equation}
\chi\nd_{\sigma\sigma}(q)={g\over q^2+m^2} + \bigg({g\over q^2+m^2}\bigg)^{\!\!2}\>\Sigma\nd_\sigma(q) + \CO(g^3)\ .
\end{equation}
For small $q$, $\Sigma\nd_\sigma(q)$ is dominated by the polarization insertion $\Pi^0(q)$, computed in the next section.
We write
\begin{equation}
\Sigma\nd_\sigma(q)=\sPi^0(q) + \ldots
\end{equation}
where $\ldots$ denotes terms that are either of higher order in $g$ or infrared finite.

\subsection{Polarization insertion}
The integral we must do is
\begin{equation}
\sPi^0(q)\equiv {m_0^4r^2(N-1)\over N} I_D(q),
\end{equation}
Here $D=d+1$, where $d$ is the spatial dimension, and
\begin{eqnarray}
I_D(q)&\equiv&\half\int\limits^\sLambda\!\!{d^Dk\over (2\pi)^D}\>{1\over\Bk^2 (\Bk+\Bq)^2}\\
&=&{q^{D-4}\over2(2\pi)^D} \!\!\!
	\int\limits_0^{\sLambda/q}\!\!\!dp\>p^{D-3}\int\!{d\sOmega\nd_D\over p^2 + 2p\cos\theta\nd_1 + 1}\ .\nonumber 
\end{eqnarray}

The $D$-dimensional unit vector is
\begin{equation}
\nhat=\big(\cos\theta\nd_1,\sin\theta\nd_1\cos\theta\nd_2,\ldots,
\sin\theta\nd_1\cdots\sin\theta\nd_{D-2}\cos\phi\big),\nonumber
\end{equation}
and the metric is
\begin{equation}
d\sOmega\nd_D=(\sin^{D-2}\!\theta\nd_1\, d\theta\nd_1) \cdots (\sin\theta\nd_{D-2} \, d\theta\nd_{D-2}) \, d\phi
\end{equation}
Here $D=d+1$, where $d$ is the spatial dimension.

For $d=2$, we have $D=3$ and we can take $\sLambda\to \infty$,
\begin{align}
	I_3(q)&=\frac{1}{2q}\frac{1}{8\pi^3}\!\int\limits_0^\infty \!\!dp\! \int\limits_0^\pi\!\!d\theta\,\sin\theta\!\!\int\limits_0^{2\pi}\!\!d\phi\>
	\frac{1}{p^2 + 2p\cos\theta + 1}\nonumber\\
	&=\frac{1}{8\pi^2 q}\int\limits_0^\infty\!\!dp\!\int\limits_{-1}^1\!\!dx\>\frac{1}{p^2 + 2xp + 1}\nonumber\\
	&=\frac{1}{8\pi^2 q}\int\limits_0^\infty \frac{dp}{p}\ln \frac{p+1}{p-1}=\frac{1}{4\pi^2 q}\int\limits_0^1 \frac{dp}{p}\ln\frac{1+p}{1-p}\nonumber\\
	&=\frac{1}{2\pi^2 q}\left( 1 + \frac{1}{3^2} + \frac{1}{5^2} + \ldots\right)= {1\over 16 \,q}\ .\nonumber
\end{align}

For $d=3$, we have $D=4$.  We must retain the ultraviolet cutoff $\sLambda$, which we take for convenience  to be isotropic in the spatial and temporal dimensions.
We make use of
\begin{equation}
\int\limits_{-\pi}^\pi \!\! d\psi\>{1\over a + b\cos\psi}={2\pi\over\sqrt{a^2-b^2}}\cdot \Theta(a^2-b^2)\ .
\end{equation}
We have
\begin{align}
I_4(q)&=\frac{1}{2}\frac{1}{(2\pi)^4}\!\int\limits_0^{\sLambda/q}\!\!\!dp\>p\int\limits_0^\pi\!\!d\theta\,\sin^2\!\theta\cdot 4\pi\cdot
\frac{1}{p^2 + 2p\cos\theta + 1}\nonumber\\
&=\frac{1}{16\pi^3}\!\!\int\limits_0^{\sLambda/q}\!\!\!dp\int\limits_{-\pi}^\pi\!\!d\theta\>\frac{p(1-\cos^2\!\theta)}{p^2 +2p\cos\theta + 1}\nonumber\\
&=\frac{1}{8\pi^2}\!\!\int\limits_0^{\sLambda/q}\!\!\!dp\left[ \frac{p}{|p^2-1| } + \frac{p^2+1}{4p} - \frac{(p^2+1)^2}{4p}\frac{1}{|p^2-1|}\right]\nonumber\\
&=\frac{1}{8\pi^2}\!\!\int\limits_0^{\sLambda/q}\!\!\!dp\left[ \frac{(p^2+1)- |p^2-1|}{4p}\right]
={1\over 32\pi^2}\left[ 1 + \ln \! \bigg({\sLambda^2\over q^2}\bigg)\right]\ .
\end{align}

These results for $d=2$ and $d=3$ are consistent with the general results in Eqs. (13) and (14) of Ref. \cite{Dupuis}.

\subsection{Analytic continuation to real frequency}
We set
\bea
q&=&\sqrt{\Bq^2-(\omega+i\eps)^2}\nonumber\\
&=& \begin{cases}\ \sqrt{\Bq^2-\omega^2} & {\rm if}\quad \omega^2 < \Bq^2 \\
e^{-i\pi/2}\sqrt{\omega^2-\Bq^2}  &  {\rm if}\quad \omega^2 > \Bq^2\ . \end{cases}
\eea
The susceptibility is
\begin{eqnarray}
\chi\nd_{\sigma\sigma}&=&{g\over\Bq^2+1-(\omega+i\eps)^2} \nonumber\\
&+& g^2{(N-1)m_0^4 r^2\over 16 N} {1 \over  \left[\Bq^2+1-(\omega+i\eps)^2\right]}\nonumber\\
&\times&
\begin{cases} {1\over\sqrt{\Bq^2-(\omega+i\eps)^2}} &  (d=2) \\
{1\over 2\pi^2}\left[ 1 + \ln \!\left( {\sLambda^2\over \Bq^2 - (\omega+i\eps)^2}\right)\right] &  (d=3) \end{cases}
\end{eqnarray}
Therefore,
\bea
\chi''_{\sigma\sigma}&=&\ {\pi g\over 2\sqrt{\Bq^2+m^2}}\,\delta\big(\omega-\sqrt{\Bq^2+m^2}\big)  \nonumber\\
&&\qquad + g^2\>{(N-1) \, m_0^4r^2\over 32\pi N} \>  {\Theta(\omega^2-\Bq^2)\over (\Bq^2+m^2-\omega^2)^2}\nonumber\\
&&\qquad\qquad\quad \times \begin{cases} {2\pi\over\sqrt{\omega^2-\Bq^2}} & (d=2) \\ \quad\  1& (d=3)\end{cases}
\eea

\section{RPA and large $N$ theory}
\label{app:RPA}
\begin{figure}[!t]
\centering
\includegraphics[width=8cm]{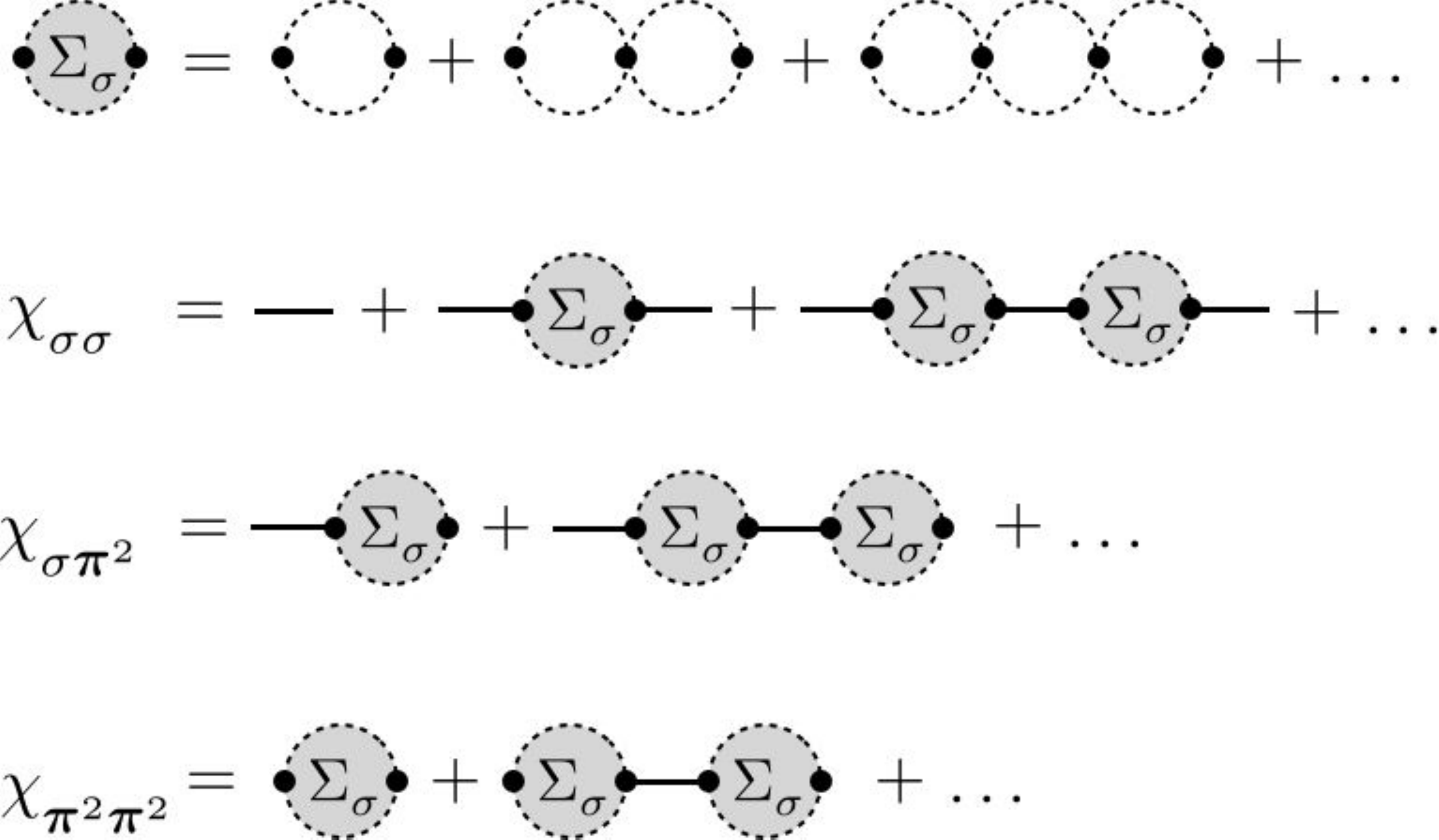}
\caption
{\label{fig:largeNsusc} Diagrams in the large-$N$ limit. For $N=\infty$, only the RPA sums shown here contribute. }
\end{figure}

Let us evaluate $\Sigma\nd_\sigma(q)$ to order $K$ in perturbation theory, where we include two $\sigma\Bpi^2$ vertices and $K$
$(\Bpi^2)^2$ vertices (see Fig.~\ref{fig:largeNsusc}).
There is an overall factor of $(-1)^{K+2}/(K+2)!$ from the exponential.  Selecting two of the $\sigma\Bpi^2$ vertices from the $K+2$
terms results in a combinatoric factor ${K+2\choose 2}$.  The two $\sigma\Bpi^2$ vertices can be interchanged, and the $K$ $(\Bpi^2)^2$ vertices
can be permuted, resulting in a factor of $2! \cdot K!$.  When we contract one of the $\Bpi$ legs of the first $\sigma\Bpi^2$ vertex with the first $(\Bpi^2)^2$ vertex, there are four choices of legs from the latter vertex to choose from.  For the second leg, there are three choices, but if we want to maximize
powers of $N$ there is only one choice.  (Recall that we are contracting $\sigma \pi^\alpha\pi^\alpha$ with $\pi^\beta\pi^\beta\pi^\gamma\pi^\gamma$.)
We are left with two uncontracted legs of the first $(\Bpi^2)^2$ vertex, and we get another factor of four from the second $(\Bpi^2)^2$ vertex.
After running through all $K$ of these $(\Bpi^2)^2$ vertices, we have two remaining legs to contract with the $\Bpi$ legs from the second $\sigma\Bpi^2$
vertex, yielding two possibilities.   Each of the $K$ $(\Bpi^2)^2$ vertices comes with a factor $m_0^2/8gN$, and each of the two $\sigma\Bpi^2$ vertices
comes with a factor $m_0^2 r/2g\sqrt{N}$.  After all the legs are contracted, we are left with $(K+1)$ loops, each containing a $\pi^\alpha$ propagator at
momentum $k+q$ and a $\pi^\alpha$ propagator at momentum $-k$.   The propagator at momentum $k$ is $g/k^2$, where $k^2=k\nd_\mu k\nd_\mu$.
For each of the loops, there are $N-1$ choices of the vector index $\alpha$ for $\pi^\alpha$.   Putting this all together, we obtain a contribution
\begin{align}
&\frac{(-1)^{K+2}}{(K+2)!}\cdot{K+2\choose 2}\cdot 2!\cdot K!\cdot 4^K\cdot 2\cdot\left(\frac{m_0^2 r}{2g\sqrt{N}}\right)^{\!\!2}\cdot\left(\frac{m_0^2}{8gN}\right)^{\!\!K}\nonumber\\
&\quad\cdot\left(\int\!\!\frac{d^D\!k}{(2\pi)^D}\>\frac{(N-1)g^2}{k^2\, (k+q)^2}\right)^{\!\!K+1}=\sPi^0(q)\cdot\left(-{g\over m^2}\sPi^0(q)\right)^{\!\!K}.\nonumber
\end{align}
Summing this over all non-negative $K$, we obtain the RPA self-energy
for the $\sigma$ field,
\begin{equation}
\Sigma^\ssr{RPA}_\sigma(q)= {\sPi^0(q)\over
1+g\,\sPi^0(q)/m^2}\ .
\end{equation}
The RPA captures the leading order behavior in the $N\to\infty$ limit.
Thus,
\begin{equation}
\chi^{N=\infty}_{\sigma\sigma}(q)={g\over q^2+m^2-g\Sigma^\ssr{RPA}_\sigma(q)}.
\end{equation}
Similarly, the scalar susceptibility is obtained from Eq.~(\ref{chiRR}) using the RPA sums shown in Fig.~\ref{fig:largeNsusc}:
\begin{equation}
\chi^{N=\infty}_{\rho\rho}(q)={4g r^2\over q^2+m^2}\,\Bigg( 1 + {{g\,q^4\,\Sigma^\ssr{RPA}_\sigma(q)/m^4}\over q^2+m^2 - g\Sigma^\ssr{RPA}_\sigma(q)}\Bigg)\ .
\end{equation}

\section{Conductivity of O(N) models}
\label{app:ONconductivity}

An $O(N)$ field theory can couple to a set of gauge, fields, with which currents and conductivities can be defined.
 Consider a local gauge transformation
\begin{equation}
\BPhi(x)\to \CO(x)\,\BPhi(x) = e^{\sTheta\nd_a(x) \,T^a}\,\BPhi(x)\ ,
\end{equation}
where $\{T^a\}$ are the $\half N(N-1)$ generators of the group $O(N)$.  The generators are real antisymmetric matrices which we normalize
according to the convention $\Tra(T^a T^b)=-2\,\delta^{ab}$.  A convenient basis then is the set of matrices
\begin{equation}
T^a_{ij}=\delta\nd_{i,I}\,\delta\nd_{j,J} - \delta\nd_{i,J}\,\delta\nd_{j,I}\ ,
\end{equation}
where $a$ denotes the composite index $(I,J)$, where $I < J$, which runs from $1$ to $\half N(N-1)$.  For this basis we have
\begin{equation}
T^a_{ij}\,T^a_{kl} = \delta\nd_{ik}\,\delta\nd_{jl} - \delta\nd_{il}\,\delta\nd_{jk}\ .
\label{ONnorm}
\end{equation}

The gauged $O(N)$ model is defined by the Lagrangian density
\begin{equation}
\CL\nd_\RE=\frac{1}{2g}\left(\pz\nd_\mu\BPhi + A\nd_\mu\BPhi\right)^2 + \frac{m_0^2}{8Ng}\left(|\BPhi|^2-N\right)^{\!2}\ ,
\end{equation}
where $A\nd_\mu$ is an antisymmetric tensor vector potential which can be expanded in the generators, {\it viz.\/}, $A\nd_\mu=A\nd_{a\mu}\,T^a$.
Gauge invariance follows from the gauge transformation rules,
\begin{align}
\BPhi&\to\CO\BPhi \\
A\nd_\mu &\to \CO^\ST A\nd_\mu \CO - \CO^\ST \pz\nd_\mu \CO\ .
\end{align}

There are $\half N(N-1)$ $O(N)$ currents, one for each generator.  We have
\begin{align}
I\nd_{a\mu}(x)&={\delta S\nd_\RE\over\delta A\nd_{a\mu}(x)} \\
&={1\over g}\,\pz\nd_\mu\BPhi\cdot T^a\BPhi + {1\over g}\,A\nd_{b\mu}\,T^a\BPhi\cdot T^b\BPhi\nonumber\\
&\equiv I^\SP_{a\mu}(x) + I^\SD_{a\mu}(x)\ ,
\end{align}
where $I^{\SP(\SD)}_{a\mu}$ is the paramagnetic (diamagnetic) current.
The corresponding Kubo formula is
\begin{equation}
\big\langle I\nd_{a\mu}(x)\big\rangle = -\int\!\!d^D\!x'\,K^{ab}_{\mu\nu}(x,x')\,A\nd_{b\nu}(x') + \CO(A^2)\ ,
\end{equation}
with
\bea
K^{ab}_{\mu\nu}(x,x')&=& \big\langle I\nd_{a\mu}(x)\,I\nd_{b\nu}(x')\big\rangle \\
&\,&- {1\over g}\,\delta\nd_{\mu\nu}\,\delta^4(x-x')\,
\big\langle T^a\BPhi(x)\cdot T^b\BPhi(x) \big\rangle\ .\nonumber
\eea
We separate $K$ into paramagnetic and diamagnetic contributions, with $K^{ab}_{\mu\nu}=K^{\SP\,ab}_{\mu\nu}+K^{\SD\,ab}_{\mu\nu}$
and $K^{\SP\,ab}_{\mu\nu}(x,x')=\langle I^\textsf{P}_{a\mu}(x)\,I^\textsf{P}_{b\nu}(x')\rangle$.

\subsection{Symmetric phase}

In the symmetric phase, $\langle \BPhi\rangle=0$, and the response function $K^{ab}_{\mu\nu}(x,x')$ is diagonal in $(a,b)$.
Summing over all $O(N)$ generator indices, we define
\beq
K\nd_{\mu\nu}(x,x')={2\over N(N-1)}\!\!\!\sum_{a=1}^{{1\over 2}N(N-1)}\!\!\! K^{aa}_{\mu\nu}(x,x'),\nonumber\\
\eeq
which equals
\begin{align}
&{1\over N(N-1)g^2}\>\Big\langle \big( \sPhi\nd_i \pz\nd_\mu\sPhi\nd_j - \sPhi\nd_j \pz\nd_\mu\sPhi\nd_i \big)\nd_x \>
\big( \sPhi\nd_i \pz\nd_\nu\sPhi\nd_j - \sPhi\nd_j \pz\nd_\nu\sPhi\nd_i \big)\nd_{x'}\Big\rangle \nonumber \\
&\quad\quad\quad - {2\over Ng}\delta\nd_{\mu\nu} \delta^4(x-x')\,\big\langle \sPhi\nd_i(x)\sPhi\nd_i(x')\big\rangle.
\label{eq:appDsymm}
\end{align}
Note that we have normalized by dividing by the total number of generators.
We can define an $O(N)$ conductivity $\sigma(\omega)$ as
\begin{equation}
\sigma(\omega)={i\over\omega d}\,\sum_{\mu=1}^d K_{\mu\mu}(\omega,\Bq=0)\ ,\label{eq:KtoSigma}
\end{equation}
where the sum is over the spatial values of the space-time indices.

In the symmetric phase, we write $\BPhi=\rho\>\nhat$ (note that this differs from the convention used in the remainder of the the text, Eq.~(\ref{parametrizations}), where we expand about the ordered phase).  We then obtain,
\begin{equation}
\pz\nd_\mu\BPhi=\rho\,\pz\nd_\mu\nhat + \nhat\,\pz\nd_\mu\rho\ .
\end{equation}
Making use of Eq. (\ref{eq:appDsymm}), we can write
\begin{align}
K\nd_{\mu\nu}(x,x')&=\frac{1}{N(N-1) g^2}\Big\langle J_{ij,\mu}(x)
\> J_{ij,\nu}(x')\Big\rangle\label{Kmunu}\\
&\quad - \frac{2}{Ng}\delta\nd_{\mu\nu} \delta^4(x-x')\,\left\langle\, \rho(x) \rho(x')\right\rangle.\nonumber
\end{align}
where
\beq
J_{ij,\mu}(x)\equiv\rho(x)\,\left( n\nd_i \pz\nd_\mu n\nd_j - n\nd_j \pz\nd_\mu n\nd_i\right)_x.
\eeq
Note that for $N=2$ we have $\nhat=(\cos\varphi \, , \,\sin\varphi)$ and $(n\nd_2\,\pz\nd_\mu n\nd_1 - n\nd_1\,\pz\nd_\mu n\nd_2)=\pz\nd_\mu\,\varphi$\,,
in which case the above expression reduces to a familiar form.

\subsection{Broken symmetry phase}

When the $O(N)$ symmetry is broken, the generators $T^a$ fall into two classes.  We define class $\SA$ generators as those which rotate between
$\sPhi^1\equiv\sigma$ and $\sPhi^{1+j}\equiv\pi\nd_j$.  There are $(N-1)$ generators of this class, with $j=1,\ldots,N-1$.  Class $\SB$ generators
rotate between $\sPhi^{1+j}$ and $\sPhi^{1+j'}$.  There are $\half (N-1)(N-2)$ generators of this class.  Note that the total number of generators
in classes $\SA$ and $\SB$ is $(N-1)+\half(N-1)(N-2)=\half N(N-1)$, the dimension of $O(N)$.  Thus, we can take
\begin{align}
T^{(j)}_{kl}&=\delta\nd_{k,1}\delta\nd_{l,j}-\delta\nd_{l,1}\delta\nd_{k,j} \quad \ \, (1<j) \\
T^{(jj')}_{kl}&=\delta\nd_{k,j}\delta\nd_{l,j'}-\delta\nd_{l,j}\delta\nd_{k,j'} \quad (1<j<j')\ .
\end{align}
The response function $K^{ab}_{\mu\nu}$ is diagonal in the generator indices, so we can in principle study two response functions,
$K^{\SA\SA}_{\mu\nu}$ and $K^{\SB\SB}_{\mu\nu}$.  Note that $K^{\SA\SB}_{\mu\nu}=K^{\SB\SA}_{\mu\nu}=0$.  

For the class $\SA$ generator $T^a$ with $a=(j)$, we have
\begin{equation}
I^\SP_{a\mu}={1\over g}\Big( \pi\nd_j \, \pz\nd_\mu\sigma - \big( r\sqrt{N}+\sigma\big)\,\pz\nd_\mu\pi\nd_j\Big) \ .
\end{equation}
The diamagnetic contribution to the response function is
\begin{align}
K^{\SD\,ab}_{\mu\nu}(x,x')&=-{1\over g}\delta\nd_{ab} \delta\nd_{\mu\nu} \delta^{(4)}(x-x') \nonumber\\
&\qquad \cdot\left\langle \big( r \sqrt{N} + \sigma(x) \big)^2 + \pi_j^2(x) \right\rangle\ .
\end{align}

For the class $\SB$ generator $T^a$ with $a=(jj')$, we have
\begin{equation}
I^\SP_{a\mu}={1\over g}\Big( \pi\nd_{j'}  \pz\nd_\mu\pi\nd_j -  \pi\nd_j \pz\nd_\mu\pi\nd_{j'}\Big)\ .
\end{equation}
The diamagnetic contribution to the response function is
\begin{equation}
K^{\SD\,ab}_{\mu\nu}(x,x')=-{1\over g}\delta\nd_{ab}  \delta\nd_{\mu\nu}  \delta^{(4)}(x-x')  
\Big\langle \pi_j^2(x) +  \pi_{j'}^2(x) \Big\rangle.
\end{equation}

Note that for $N=2$, class $\SB$ is the empty set.  Hence, in analogy with the $O(2)$ conductivity, in what follows we focus on the class $\SA$ response function and will drop the $\SA\SA$ superscript.  Then, averaging over the $(N-1)$ generators in this class, we have
\begin{align}
K^{\SP}_{\mu\nu}&(x,x')=\frac{ r^2N}{(N-1) g^2}\Big\langle \pz\nd_\mu \Bpi\nd_x \cdot  \pz\nd_\nu \Bpi\nd_{x'} \Big\rangle \label{KP} \\
&+\frac{r\sqrt{N}}{(N-1)  g^2} \ \Big\langle \pz\nd_\mu \Bpi\nd_x \cdot \big( \sigma\pz\nd_\nu\Bpi - \Bpi\pz\nd_\nu\sigma\big)\nd_{x'}\Big\rangle\nonumber\\
&+\frac{r\sqrt{N}}{(N-1)  g^2} \ \Big\langle \big( \sigma\pz\nd_\mu\Bpi - \Bpi\pz\nd_\mu\sigma\big)\nd_x \cdot \pz\nd_\nu \Bpi\nd_{x'} \Big\rangle
\bvph\nonumber\\
& + \frac{1}{(N-1)  g^2}  \Big\langle \big(\sigma\pz\nd_\mu\Bpi - \Bpi\pz\nd_\mu\sigma\big)_x \cdot
\big(\sigma\pz\nd_\nu\Bpi - \Bpi\pz\nd_\nu\sigma\big)\nd_{x'}\Big\rangle.\nonumber
\end{align}

We are interested in the imaginary conductivity at $\Bq=0$ and finite frequency, which we compute up to two-loop order.
Thus we can omit the diamagnetic term, and the  first three terms of Eq.~(\ref{KP}).  Taking the Fourier transform, we then obtain
\begin{eqnarray}
K_{\mu\nu}(q)&=&{1\over (N-1)  g^2} \!\int\!\!d^D\!x\> e^{iq\cdot (x-x')} \label{eq:ONcond}\\
&\,&\times\quad\Big\langle \big(\sigma\pz\nd_\mu\Bpi - \Bpi\pz\nd_\mu\sigma\big)\nd_x \cdot
\big(\sigma\pz\nd_\nu\Bpi - \Bpi\pz\nd_\nu\sigma\big)\nd_{x'}\Big\rangle,\nonumber
\end{eqnarray}
where $q\cdot x \equiv q\nd_\mu  x^\mu$.  In what follows, we valuate this expression to one- and two-loop order and use Eq.~(\ref{eq:KtoSigma}) to compute the optical conductivity.

\section{Optical conductivity to order $g$}
\label{app:condCalculation}

\subsection{Conductivity at order $g^{0}$}

At order $g^0$, Eq.~(\ref{eq:ONcond}) factorizes into the product of a $\sigma$ propagator and a $\Bpi$ propagator.  This yields the one-loop integral,
\begin{equation}
K\nd_{\mu\nu}(q)={1\over\beta}\sum_{\nu\nd_m}\int\!\!{d^d\!k\over (2\pi)^d}\> {(2 k^\mu+q^\mu) (2 k^\nu+q^\nu)\over (k^2+m^2)(k+q)^2}\ .
\end{equation}
We set $\mu=\nu=x$ and $q=(i\omega\nd_n,\Bq=0)$.  We then have
\begin{align}
K\nd_{xx}(&i\omega\nd_n)=\frac{4}{\beta}\sum_{\nu\nd_m} \!\!
\int\!\!\frac{d^d\!k}{(2\pi)^d}\>\frac{\Bk^2}{\Bk^2+m^2+(i\nu\nd_m)^2} \frac{1}{\Bk^2 + (i\nu\nd_m+i\omega\nd_n)^2}\nonumber\\
&=\frac{4}{d}\frac{S\nd_d}{(2\pi)^d} \int\limits_0^\Lambda\!\!dk\> k^{d+1}\> \frac{1}{\beta}\sum_{\nu\nd_m}\frac{1}{i\nu\nd_m+a\nd_k}\frac{1}{i\nu\nd_m-a\nd_k}\times \label{eq:g0K}\\
&\qquad\qquad\qquad\times\frac{1}{i\nu\nd_m+i\omega\nd_n +b\nd_k}\frac{1}{i\nu\nd_m+i\omega\nd_n-b\nd_k}, \nonumber
\end{align}
where $a\nd_k=\sqrt{k^2+m^2}$ and $b\nd_k=k$.

Now the bosonic Matsubara sum can be written as
\begin{equation}
{1\over\beta}\sum_{\nu\nd_m}h(i\nu\nd_m)=-\sum_{{\tilde\nu}} n({\tilde\nu})\textsf{Res} \big[ h({\tilde\nu}) \big],\label{eq:Matsubara}
\end{equation}
where the sum is over the poles ${\tilde\nu}$ of $h$, and where $n(\nu)=\big[\exp(\nu/T)-1  \big]^{-1}$ is the Bose function.  Note that $n(\pm b - i\omega\nd_n)=n(\pm b)$ when $\omega\nd_n$ is
a bosonic Matsbara frequency.
Hence,
\begin{align}
F(a,b,i\omega\nd_n)&\equiv\frac{1}{\beta}\sum_{\nu\nd_m}\frac{1}{(i\nu\nd_m+a)(i\nu\nd_m-a)(i\nu\nd_m+i\omega\nd_n +b)(i\nu\nd_m+i\omega\nd_n-b) } \nonumber \\
&=\frac{ n(-a)}{2a \big[ (i\omega\nd_n - a)^2 - b^2 \big] } - \frac{ n(a)}{2a \big[ (i\omega\nd_n + a)^2 - b^2 \big] } \bvph \label{eq:Fres} \\
&\hskip0.2in + \frac{ n(-b)}{2b \big[ (i\omega\nd_n + b)^2 - a^2 \big] } - \frac{ n(b)}{2b \big[ (i\omega\nd_n - b)^2 - a^2 \big] }. \bvph \nonumber
\end{align}
Thus, at $T=0$,
\begin{equation}
F(a\nd_k,b\nd_k,\omega+i\eps)=\frac{1}{4 a\nd_k b\nd_k} \left[ \frac{1}{\omega + i\eps + a\nd_k + b\nd_k} - \frac{1}{\omega  + i\eps- a\nd_k - b\nd_k} \right], \label{eq:FT0}
\end{equation}
and the conductivity at order $g^0$ is
\begin{align}
\sigma(\omega)&={1\over\omega} \> \Imp K\nd_{xx}(\omega+i\eps,\Bq=0) \\
&= \frac{\pi}{d\omega}\frac{S\nd_d}{(2\pi)^d} \int\limits_0^\Lambda\!\!dk\> \frac{k^d}{\sqrt{k^2+m^2}} \> \bigg[  \delta\!\left(\omega-k-\sqrt{k^2+m^2}\right)
\nonumber\\
&\hskip1.4in - \delta\!\left(\omega+k+\sqrt{k^2+m^2}\right)\bigg] \nonumber \\
&= {\pi S\nd_d\over d\omega^2} \bigg( {\omega^2-m^2\over 4\pi\omega}\bigg)^{\!d} \>\Theta(\omega^2-m^2)\ .\nonumber
\end{align}

\begin{figure}[!t]
\centering
\includegraphics[width=8cm]{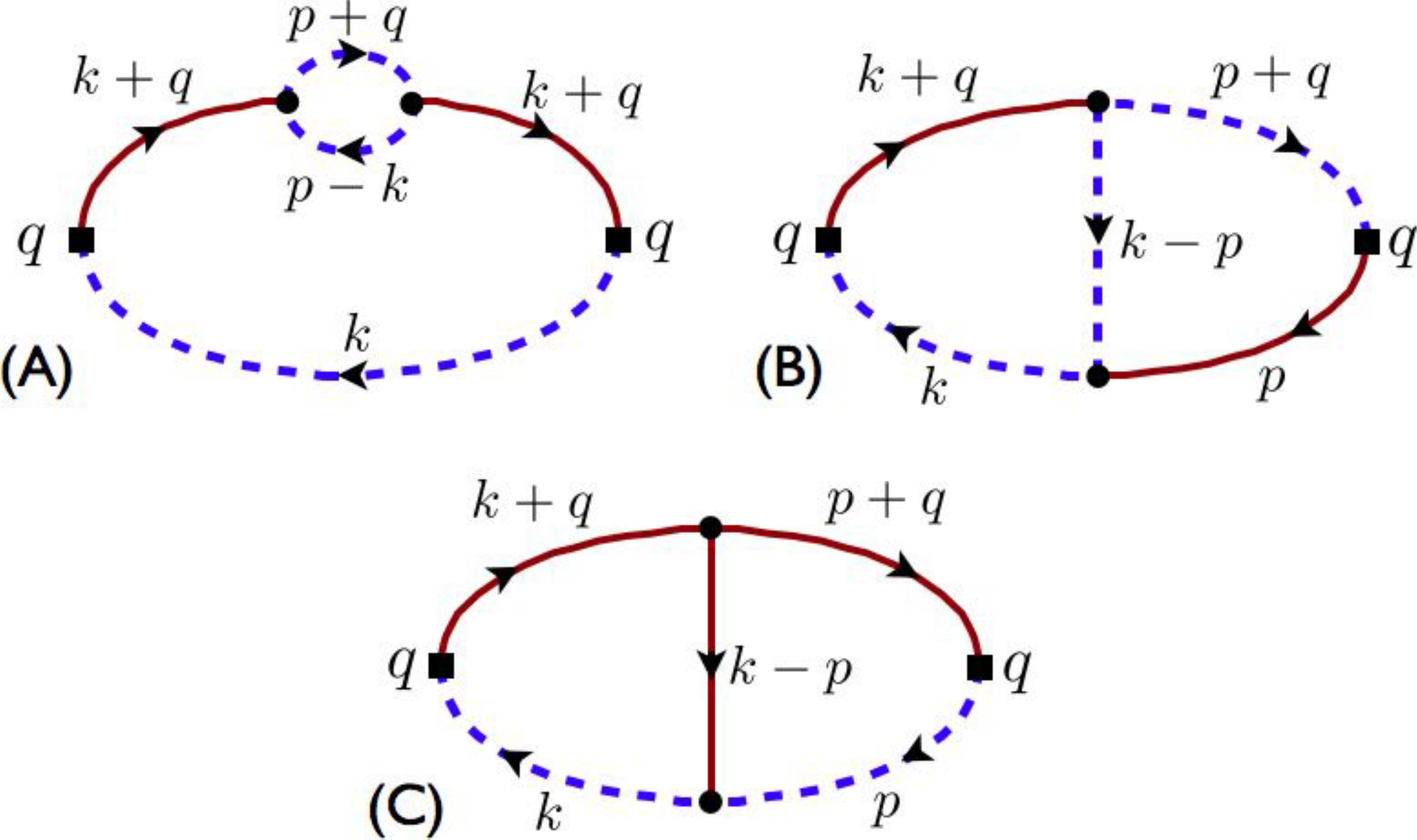}
\caption
{\label{orderg} Order $g$ diagrams contributing to the optical conductivity for $\omega < 2m$.  Solid brown lines are $\sigma$ propagators while
dashed blue lines are $\Bpi$ propagators.  Note that diagram (c) does not contribute to the subgap conductivity and is therefore not shown in Fig.~\ref{fig:sigma-diagrams}}
\end{figure}

\subsection{Conductivity at order $g$}

There are 11 two-loop diagrams which enter the conductivity at order $g$.  Of these, most either renormalize the zero-frequency superfluid stiffness peak or have a threshold at high frequency.  This leaves three contributions, diagrammatically represented in Fig.~\ref{orderg}, which affect the finite frequency response below $\omega=2m$.  We have
\begin{equation}
q=(0\, , \, i\omega\nd_j),\quad k = (\Bk \, , \, i\nu\nd_m),\quad p=(\Bp \, , \, i\xi\nd_n)\ .
\end{equation}
We find it convenient to define
\begin{equation}
\Bk\nd_1 = \Bk,\quad \Bk\nd_2=-\Bp,\quad \Bk\nd_3=\Bp-\Bk,
\end{equation}
so that $\Bk\nd_1+\Bk\nd_2 +\Bk\nd_3=0$, \ie\ the vectors $\Bk\nd_{1,2,3}$ form the legs of a triangle.

The contributions $K\nd_\RA$, $K\nd_\RB$, and $K\nd_\RC$ are given by
\begin{align}
K\nd_\RA&=\frac{1}{(N-1)g^2} \cdot \frac{m^4}{8 N g^2}\cdot (N-1)^2\cdot 4\cdot \frac{4}{d}	\int\!\!\frac{d^d\!k\nd_1}{(2\pi)^d}\!\!
	\int\!\!\frac{d^d\!k\nd_2}{(2\pi)^d}\nonumber \\
&\quad \cdot \frac{1}{\beta}\sum_{\nu\nd_m}\ \frac{1}{\beta}\sum_{\xi\nd_n} k_1^2\, \frac{g}{\left[k_1^2+m^2-(i\nu\nd_m+i\omega\nd_j)^2\right]^2}\frac{g}{k_1^2-(i\nu\nd_m)^2} \nonumber\\
&\qquad\cdot \frac{g}{k_2^2-(i\xi\nd_n+i\omega\nd_j)^2} \frac{g}{k_3^2-(i\xi\nd_n-i\nu\nd_m)^2}, 
\end{align}
\begin{align}
K\nd_\RB&=-\frac{1}{(N-1)g^2}\cdot \frac{m^4}{8 N g^2}\cdot (N-1)\cdot 8\cdot \frac{4}{d}	\! \int\!\!\frac{d^d\!k\nd_1}{(2\pi)^d}\!\!
	\int\!\!\frac{d^d\!k\nd_2}{(2\pi)^d}\nonumber\\
&\quad\cdot\ \frac{1}{\beta}\sum_{\nu\nd_m}\ \frac{1}{\beta}\sum_{\xi\nd_n} \Bk\nd_1\cdot\Bk\nd_2 \frac{g}{k_1^2+m^2-(i\nu\nd_m+i\omega\nd_j)^2}\frac{g}{k_2^2+m^2-(i\xi\nd_n)^2}\nonumber\\
&\qquad \cdot \frac{g}{k_1^2-(i\nu\nd_m)^2} \cdot \frac{g}{k_3^2-(i\nu\nd_m-i\xi\nd_n)^2} \frac{g}{k_2^2-(i\xi\nd_n)^2},
\end{align}
and
\begin{align}
K\nd_\RC&=\frac{1}{(N-1)g^2}\cdot \frac{m^4}{8 N g^2}\cdot (N-1)\cdot 12\cdot \frac{4}{d}	\! \int\!\!\frac{d^d\!k\nd_1}{(2\pi)^d}\!\!
	\int\!\!\frac{d^d\!k\nd_2}{(2\pi)^d}\nonumber\\
&\quad\cdot\ \frac{1}{\beta}\sum_{\nu\nd_m}\ \frac{1}{\beta}\sum_{\xi\nd_n} \Bk\nd_1\cdot\Bk\nd_2\frac{g}{k_1^2-(i\nu\nd_m)^2} \frac{g}{k_2^2+m^2-(i\omega\nd_j+i\xi\nd_n)^2}\nonumber\\
&\qquad \cdot
\frac{g}{k_3^2+m^2-(i\nu\nd_m-i\xi\nd_n)^2} \frac{g}{k_1^2+m^2-(i\nu\nd_m+i\omega\nd_j)^2}\frac{g}{k_2^2-(i\xi\nd_n)^2}.  
\end{align}

\begin{widetext}

Carrying out the Matsubara sum using Eq.~(\ref{eq:Matsubara}), first over $\xi\nd_n$ and then over $\nu\nd_m$, we obtain
\begin{align}
K\nd_\RA&=-\frac{gm^4\,(N-1)}{ Nd}\!\!  \int\!\!{d^d\!k\nd_1\over (2\pi)^d}\!\! \int\!\!{d^d\!k\nd_2\over (2\pi)^d}
\sum_\nu\textsf{Res}\bigg[\frac{k_1^2}{k\nd_2} \frac{n(\nu)}{ (\nu+i\omega\nd_j+k\nd_2)^2-k_3^2}  \frac{1}{\nu^2-k_1^2} 
\frac{1}{\left[ (\nu+i\omega\nd_j)^2-k_1^2-m^2\right]^2} \nonumber\\
&\hskip1.7in+ \frac{k_1^2}{ k\nd_3} \frac{n(\nu)}{(\nu+i\omega\nd_j-k\nd_3)^2-k_2^2}  \frac{1}{\nu^2-k_1^2} 
\frac{1}{\left[ (\nu+i\omega\nd_j)^2-k_1^2-m^2\right]^2}\bigg],
\end{align}
\begin{align}
K\nd_\RB&=\frac{gm^4}{Nd}\!\!  \int\!\!\frac{d^d\!k\nd_1}{(2\pi)^d}\!\! \int\!\!\frac{d^d\!k\nd_2}{(2\pi)^d} \> (k_1^2+k_2^2-k_3^2) \sum_\nu\textsf{Res} \bigg[ \frac{n(\nu)}{\sqrt{k_2^2+m^2}}  \frac{1}{ \left(i\omega\nd_j-\sqrt{k_2^2+m^2}\right)^2-k_2^2} 
\frac{1}{\left(\nu+\sqrt{k_2^2+m^2}\right)-k_3^2}  \frac{1}{\nu^2 - k_1^2}  \frac{1}{(\nu+i\omega\nd_j)^2-k_1^2-m^2}\nonumber\\
&\qquad\qquad+ \frac{n(\nu)}{k\nd_2}  \frac{1}{(i\omega\nd_j+k\nd_2)^2-k_2^2-m^2} 
\frac{1}{(\nu+i\omega\nd_j+k\nd_2)^2-k_3^2}  \frac{1}{\nu^2-k_1^2}  \frac{1}{(\nu+i\omega\nd_j)^2-k_1^2-m^2}\nonumber\\
&\qquad\qquad + \frac{n(\nu)}{k\nd_3}  \frac{1}{(\nu-k\nd_3)^2-k_2^2}  \frac{1}{(\nu+i\omega\nd_j-k\nd_3)^2-k_2^2} 
\frac{1}{\nu^2-k_1^2}  \frac{1}{(\nu+i\omega\nd_j)^2-k_1^2-m^2} \bigg]\nonumber,
\end{align}
and
\begin{align}
K\nd_\RC&=\frac{3gm^4}{Nd}\!\!   \int\!\!\frac{d^d\!k\nd_1}{(2\pi)^d}\!\! \int\!\!\frac{d^d\!k\nd_2}{(2\pi)^d}  \> (k_1^2+k_2^2-k_3^2) \sum_\nu
\textsf{Res} \Bigg\{ \frac{n(\nu)}{\nu^2-k_1^2}  \frac{1} { (\nu+i\omega\nd_j)^2 - k_1^2 - m^2 } \, \bigg[
\frac{1}{k\nd_2}  \frac{1}{\left(\nu+k\nd_2\right)^{\!2} - k_3^2-m^2}  \frac{1}{\left(i\omega\nd_j-k\nd_2\right)^2-k_2^2-m^2}
\nonumber\\
&\hskip 1.0in  + \frac{1}{\sqrt{k_2^2+m^2}}  \frac{1}{\left(\nu+i\omega\nd_j+\sqrt{k_2^2+m^2}\right)^{\!2} - k_3^2-m^2} 
\frac{1}{\left(i\omega\nd_j+ \sqrt{k_2^2+m^2}\right)^{\!2}-k_2^2} \nonumber\\
&\hskip 1.0in+ \frac{1}{\sqrt{k_3^2+m^2}}  \frac{1}{\left(\nu-\sqrt{k_3^2+m^2}\right)^{\!2} - k_2^2} 
\frac{1}{\left(\nu + i\omega\nd_j - \sqrt{k_3^2+m^2}\right)^{\!2}-k_2^2-m^2} \bigg] \Bigg\}\nonumber
\end{align}
Using Eq.~(\ref{eq:KtoSigma}), we next obtain the $\CO(g)$ conductivity from these expressions.

\subsubsection{The $\CO(g)$ contribution $\sigma_\RA(\omega)$}
In computing the contribution $\sigma\nd_\RA(\omega)$, we must compute
\begin{align}
\sum_\nu \textsf{Res}\bigg[ {F(\nu)\over \big[ (\nu+i\omega\nd_j)^2-b_1^2\big]^2}\bigg] &= \sum_\nu \textsf{Res}\bigg[ 
{F(\nu)\over(\nu+i\omega\nd_j+b\nd_1)^2 \, (\nu+i\omega\nd_j-b\nd_1)^2 }\bigg]\\
&={1\over 4b_1^3}\bigg[ F(-i\omega\nd_j-b\nd_1) + b\nd_1\, F'(-i\omega\nd_j-b\nd_1)- 
 F(-i\omega\nd_j+b\nd_1) + b\nd_1\,F'(-i\omega\nd_j+b\nd_1)^2 \bigg]\nonumber \\
&\qquad + \sum_\nu {1\over \big[(\nu+i\omega\nd_j)^2-b_1^2\big]^2}\cdot\textsf{Res}\big[ F(\nu) \big]\ ,\nonumber
\end{align}
where $b\nd_j=\sqrt{k_j^2+m^2}$, with $j\in\{1,2,3\}$.  Note that at $T=0$, $n(-i\omega\nd_j-b\nd_1)=-1$ and $n(-i\omega\nd_j+b\nd_1)=0$,
where $n(z)=\left(\exp(z)-1\right)^{-1}$ is the Bose function.  We now set $i\omega\nd_n=\omega+i0^+$, in which case
\begin{align}
\sigma\nd_\RA(\omega)&=\frac{gm^4\,(N-1)}{ Nd} \> \Imp \frac{1}{\omega} \!\! \int\!\!{d^d\!k\nd_1\over (2\pi)^d}\!\!
\int\!\!{d^d\!k\nd_2\over (2\pi)^d} \Bigg\{ \frac{k_1^2}{4b_1^3 k\nd_2 k\nd_3}\,
\bigg( \frac{1}{(\omega+b\nd_1)^2-k_1^2} + \frac{2b\nd_1(\omega+b\nd_1)}{\left[(\omega+b\nd_1)^2-k_1^2\right]^2}\bigg)
\,\bigg(\frac{k_3}{(b\nd_1-k\nd_2)^2-k_3^2} +\frac{k_2}{(b\nd_1+k\nd_3)^2-k_2^2} \bigg) \nonumber\\
&\hskip1.6in+ \frac{k_1^2}{4b_1^2k\nd_2k\nd_3} \, \frac{1}{(\omega+b\nd_1)^2-k_1^2} \,
\bigg( \frac{2(b\nd_1 - k\nd_2)k\nd_3}{\left[(b\nd_1-k\nd_2)^2-k_3^2\right]^2} +
\frac{2(b\nd_1 + k\nd_3)k\nd_2}{\left[(b\nd_1+k\nd_3)^2-k_2^2\right]^2}\bigg) \nonumber \\ 
&\hskip1.6in -\frac{k\nd_1}{2k\nd_2 k\nd_3}\,\frac{1}{\left[ (\omega-k\nd_1)^2-b_1^2\right]^2} \bigg( \frac{k_3}{(\omega-k\nd_1+k\nd_2)^2-k_3^2}
+ \frac{k_2}{(\omega-k\nd_1-k\nd_3)^2-k_2^2}\bigg)\nonumber\\
&\hskip1.6in- \frac{k_1^2}{2k\nd_2 k\nd_3} \, \frac{1}{ \left[ (k\nd_2+k\nd_3)^2-b_1^2 \right]^2 } \, \frac{1}{(\omega+k\nd_2+k\nd_3)^2-k_1^2}\Bigg\}\ .
\end{align} 
Combining terms, we find
\begin{align}
\sigma\nd_\RA(\omega)&=\frac{gm^4\,(N-1)}{ Nd} \> \Imp \frac{1}{\omega} \!\! \int\!\!{d^d\!k\nd_1\over (2\pi)^d}\!\!
\int\!\!{d^d\!k\nd_2\over (2\pi)^d} \Bigg\{ \frac{k_1^2}{4b_1^3 k\nd_2 k\nd_3}\,\frac{k\nd_2+k\nd_3}{b_1^2-(k\nd_2+k\nd_3)^2}
\bigg( \frac{1}{(\omega+b\nd_1)^2-k_1^2} + \frac{2b\nd_1(\omega+b\nd_1)}{\left[(\omega+b\nd_1)^2-k_1^2\right]^2}\bigg) \nonumber\\
&\hskip1.3in+ \frac{k_1^2}{2b\nd_1k\nd_2k\nd_3} \, \frac{1}{(\omega+b\nd_1)^2-k_1^2} \, \frac{k\nd_2+k\nd_3}
{\left[b_1^2-(k\nd_2+k\nd_3)^2\right]^2}-\frac{k\nd_1}{2k\nd_2 k\nd_3}\,\frac{1}{\left[ (\omega-k\nd_1)^2-b_1^2\right]^2} \,
\frac{k\nd_2+k\nd_3}{(\omega-k\nd_1)^2-(k\nd_2+k\nd_3)^2} \nonumber\\
&\hskip1.6in- \frac{k_1^2}{2k\nd_2 k\nd_3} \, \frac{1}{ \left[ (k\nd_2+k\nd_3)^2-b_1^2 \right]^2 } \, \frac{1}{(\omega+k\nd_2+k\nd_3)^2-k_1^2}\Bigg\}\ .
\end{align} 
Now we use
\begin{equation}
\Imp\>{1\over \left[(\omega-k)^2-b^2\right]^2} = {\pi\over 4b^2}\left[ \delta'(\omega-b-k) + \delta'(\omega+b-k)\right]
\end{equation}
and set $\omega>0$ to obtain
\begin{align}
\sigma\nd_\RA(\omega>0)&={\pi gm^4\,(N-1) \over 8Nd\omega} \!\! \int\!\!{d^d\!k\nd_1\over (2\pi)^d}\!\! \int\!\!{d^d\!k\nd_2\over (2\pi)^d}\,\left\{
{2k\nd_1\over k\nd_2 k\nd_3}\, {\delta(\omega-k\nd_1-k\nd_2-k\nd_3)\over \left[ b_1^2 - (k\nd_2+k\nd_3)^2\right]^2} 
- {k\nd_1 (k\nd_2+k\nd_3)\over b_1^2 \, k\nd_2 k\nd_3} \, {\delta'\!(\omega-k\nd_1-b\nd_1) \over
b_1^2 - (k\nd_2+k\nd_3)^2}\right\} \ . \label{dsigA}
\end{align}
The first term inside the large braces integrates to a result which is finite for arbitrarily small $\omega$.  The second term yields a threshold behavior
and is finite only for $\omega\ge m$.

\subsubsection{The $\CO(g)$ contribution $\sigma_\RB$}
Summing the residues, we find
\begin{align}
\sigma\nd_\RB(\omega)&=\frac{gm^4}{ 2Nd} \> \Imp \frac{1}{\omega} \!\!\int\!\!{d^d\!k\nd_1\over (2\pi)^d}\!\!
\int\!\!{d^d\!k\nd_2\over (2\pi)^d} \ \big(k_1^2+k_2^2-k_3^2\big) \> \bigg\{\frac{1}{k\nd_1b\nd_2} \, \frac{1}{(\omega-b\nd_2)^2-k_2^2} \,
\frac{1}{(b\nd_2-k\nd_1)^2-k_3^2} \, \frac{1}{(\omega-k\nd_1)^2 - b_1^2} \nonumber\\
&\qquad + \frac{1}{k\nd_1 k\nd_2} \, \frac{1}{(\omega+k\nd_2)^2-b_2^2} \, \frac{1}{(\omega-k\nd_1+k\nd_2)^2-k_3^2} \, 
\frac{1}{(\omega-k\nd_1)^2 - b_1^2} + \frac{1}{k\nd_1 k\nd_3} \, \frac{1}{(k\nd_1+k\nd_3)^2-b_2^2} \, \frac{1}{(\omega-k\nd_1-k\nd_3)^2-k_2^2} \, 
\frac{1}{(\omega-k\nd_1)^2 - b_1^2}\nonumber\\
&\qquad + \frac{1}{b\nd_1 b\nd_2} \, \frac{1}{(\omega-b\nd_2)^2-k_2^2} \, \frac{1}{(\omega+b\nd_1-b\nd_2)^2-k_3^2} \, 
\frac{1}{(\omega+b\nd_1)^2 - k_1^2} + \frac{1}{b\nd_1 k\nd_2} \, \frac{1}{(\omega+k\nd_2)^2-b_2^2} \, \frac{1}{(b\nd_1-k\nd_2)^2-k_3^2} \, 
\frac{1}{(\omega+b\nd_1)^2 - k_1^2}\nonumber\\
&\qquad + \frac{1}{b\nd_1 k\nd_3} \, \frac{1}{(\omega+b\nd_1+k\nd_3)^2-b_2^2} \, \frac{1}{(b\nd_1+k\nd_3)^2-k_2^2} \, 
\frac{1}{(\omega+b\nd_1)^2 - k_1^2} + \frac{1}{b\nd_2 k\nd_3} \, \frac{1}{(\omega-b\nd_2)^2-k_2^2} \, \frac{1}{(b\nd_2+k\nd_3)^2-k_1^2} \, 
\frac{1}{(\omega-b\nd_2-k\nd_3)^2 - b_1^2}\nonumber\\
&\qquad + \frac{1}{k\nd_2 k\nd_3} \, \frac{1}{(\omega+k\nd_2)^2-b_2^2} \, \frac{1}{(\omega+k\nd_2+k\nd_3)^2-k_1^2} \, 
\frac{1}{(k\nd_2+k\nd_3)^2 - b_1^2}\bigg\} \ .
\end{align} 
After some grinding, we obtain
\begin{align}
\sigma\nd_\RB(\omega>0)&=-\frac{\pi gm^4}{ 4Nd\omega} \!\int\!\!{d^d\!k\nd_1\over (2\pi)^d}\!\!
\int\!\!{d^d\!k\nd_2\over (2\pi)^d} \ \big(k_1^2+k_2^2-k_3^2\big) \> \bigg\{ \Big[
\frac{1}{b\nd_1 k\nd_1 k\nd_2} \, \frac{1}{(b\nd_1+k\nd_2)^2-k_3^2} \, \frac{1}{(b\nd_1+k\nd_1+k\nd_2)^2-b_2^2}\nonumber\\
&\hskip0.3in + \frac{1}{b\nd_1 k\nd_1 k\nd_2} \, \frac{1}{(b\nd_1-k\nd_2)^2-k_3^2} \, \frac{1}{(b\nd_1+k\nd_1-k\nd_2)^2-b_2^2}
+\frac{1}{k\nd_1 b\nd_1 b\nd_2} \, \frac{1}{(b\nd_2-k\nd_1)^2-k_3^2} \, \frac{1}{(k\nd_1+b\nd_1-b\nd_2)^2-k_2^2} \nonumber\\
&\hskip0.3in +\frac{1}{k\nd_1 b\nd_1 b\nd_2} \, \frac{1}{(b\nd_2+k\nd_1)^2-k_3^2} \, \frac{1}{(b\nd_1+b\nd_2+k\nd_2)^2-k_1^2} 
+\frac{1}{k\nd_1 b\nd_1 k\nd_3} \, \frac{1}{(k\nd_1+k\nd_3)^2-b_2^2} \, \frac{1}{(b\nd_1-k\nd_3)^2-k_2^2} \nonumber\\
&\hskip0.3in +\frac{1}{k\nd_1 b\nd_1 k\nd_3} \, \frac{1}{(k\nd_1-k\nd_3)^2-b_2^2} \, \frac{1}{(b\nd_1+k\nd_3)^2-k_2^2}\Big]\,
\delta(\omega-b\nd_1-k\nd_1)\nonumber\\
&\hskip0.5in +\frac{1}{b\nd_1 b\nd_2 k\nd_3} \, \frac{1}{(b\nd_1+k\nd_3)^2-k_2^2} \, \frac{1}{(b\nd_2+k\nd_3)^2-k_1^2} \,
\delta(\omega-b\nd_1-b\nd_2-k\nd_3) \nonumber\\
&\hskip0.5in +\frac{1}{k\nd_1 k\nd_2 k\nd_3} \, \frac{1}{(k\nd_1+k\nd_3)^2-b_2^2} \, \frac{1}{(k\nd_2+k\nd_3)^2-b_1^2} \,
\delta(\omega-k\nd_1-k\nd_2-k\nd_3) \bigg\}\ . \label{dsigB}
\end{align} 
The last term inside the large braces integrates to a result which is finite for arbitrarily small $\omega$.  The first six terms, which are multiplied 
by $\delta(\omega-b\nd_1-k\nd_1)$, are finite for $\omega\ge m$.  The seventh term has a threshold $\omega=2m$.

\subsubsection{The $\CO(g)$ contribution $\sigma_\RC$}
Proceeding in a similar manner, we obtain
\begin{align}
\sigma\nd_\RC(\omega>0)&=\frac{3\pi gm^4}{ 2Nd\omega} \!\int\!\!{d^d\!k\nd_1\over (2\pi)^d}\!\!
\int\!\!{d^d\!k\nd_2\over (2\pi)^d} \ \big(k_1^2+k_2^2-k_3^2\big) \> \bigg\{ \Big[
\frac{1}{b\nd_1 k\nd_1 k\nd_2} \, \frac{1}{(k\nd_1-k\nd_2)^2-k_3^2} \, \frac{1}{(b\nd_1+k\nd_1-k\nd_2)^2-b_2^2} \nonumber \\
&\hskip0.7in + \frac{1}{b\nd_1 k\nd_1 b\nd_2} \, \frac{1}{(b\nd_1+b\nd_2)^2-b_3^2} \, \frac{1}{(b\nd_1+k\nd_1+b\nd_2)^2-k_2^2}
+\frac{1}{b\nd_1 k\nd_1 b\nd_3} \, \frac{1}{(k\nd_1+b\nd_3)^2-k_2^2} \, \frac{1}{(b\nd_1-b\nd_3)^2-b_2^2} \Big]\,
\delta(\omega-b\nd_1-k\nd_1)\nonumber\\
&\hskip1.0in +\frac{1}{k\nd_1 b\nd_2 b\nd_3} \, \frac{1}{(k\nd_1+b\nd_3)^2-k_2^2} \, \frac{1}{(b\nd_2+b\nd_3)^2-b_1^2} 
\ \delta(\omega-k\nd_1-b\nd_2-b\nd_3) \bigg\}\ . \label{dsigC}
\end{align} 
The first Dirac delta has a threshold at $\omega=m$, and the second Dirac delta has a threshold at $\omega=2m$.

\end{widetext}

\subsection{subgap conductivity to order $g$}

The complicated expressions derived above can, in principle, be evaluated numerically.  Here we focus on the limit $\omega<m$, where we can obtain an analytic expression for the subgap conductivity as a power series in $\omega/m$.  Of all the terms computed above, only the first term in $\sigma\nd_\RA$ and the last term in $\sigma\nd_\RB$ are nonzero for $\omega<m$.  These add up to
\begin{align}
\sigma\nd_g(\omega)&=\frac{\pi g m^4}{4N d} \int\!\!\frac{d^d\!k\nd_1}{(2\pi)^d}\int\!\!\frac{d^d\!k\nd_2}{(2\pi)^d} \frac{\delta(\omega-k\nd_1-k\nd_2-k\nd_3)}{\omega k\nd_1 k\nd_2 k\nd_3}\frac{1}{m^2 + k_1^2 - (k\nd_2 + k\nd_3)^2 } \bvph\times \nonumber \\
&\hskip 0.2in \times \left[ \frac{(N-1)\,k_1^2}{m^2 + k_1^2 - (k\nd_2 + k\nd_3)^2 } - \frac{k_1^2 + k_2^2 - k_3^2}{m^2+ k_2^2 - (k\nd_1 + k\nd_3)^2 } \right]. \label{OGB}
\end{align}

\begin{figure}[!t]
\centering
\includegraphics[width=6cm]{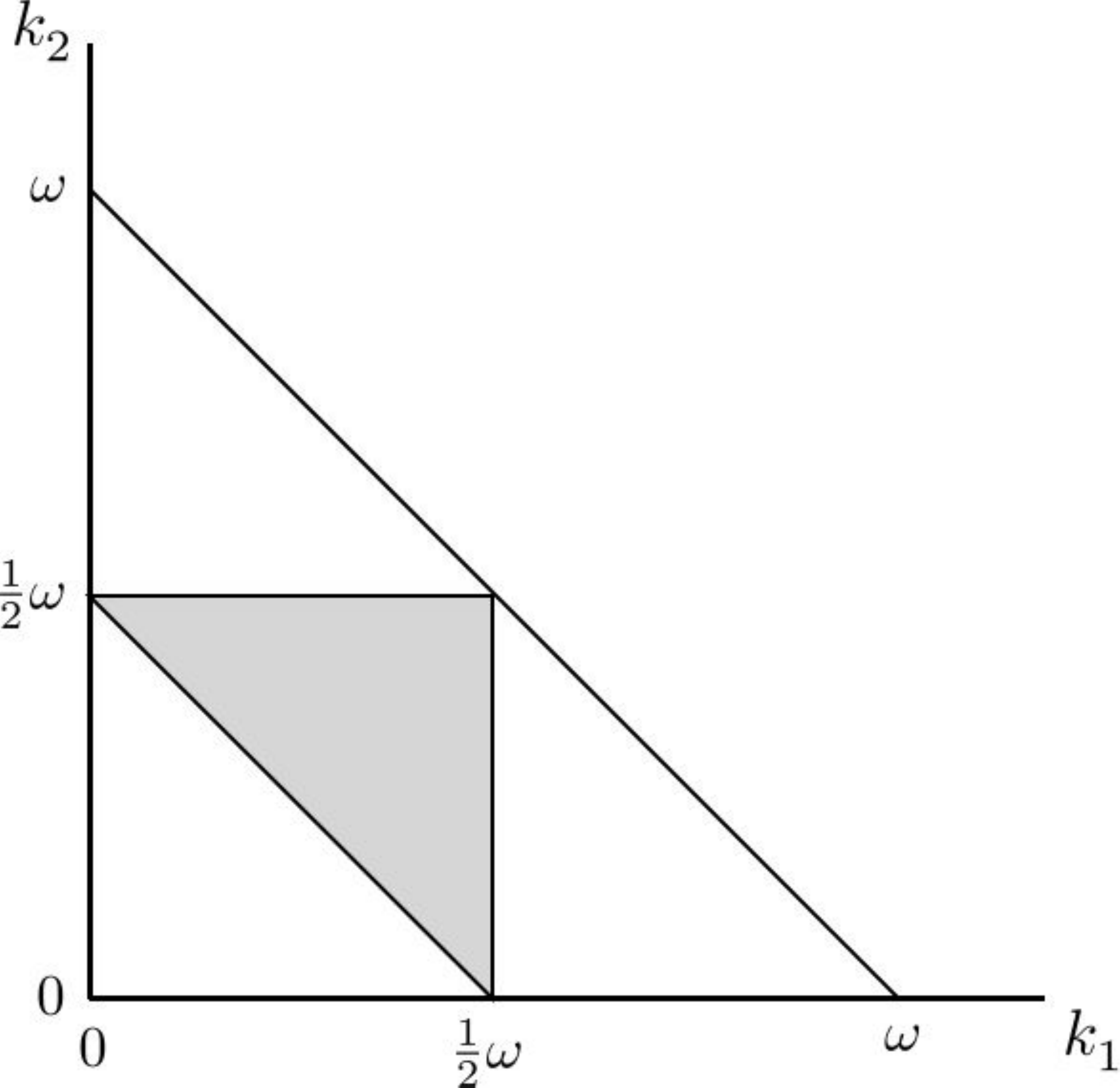}
\caption
{\label{irange} The range of $(k_1,k_2)$ integration is the gray triangle.}
\end{figure}

We can write
\begin{equation}
k\nd_3=\sqrt{k_1^2 + k_2^2 + 2xk\nd_1 k\nd_2}\ ,
\end{equation}
where $x$ is the cosine of the angle between $\Bk\nd_1$ and $\Bk\nd_2$.  Then
\begin{equation}
{\delta(\omega-k\nd_1-k\nd_2-k\nd_3)\over \omega k\nd_1 k\nd_2 k\nd_3} = {\delta\big(x-x(k\nd_1,k\nd_2,\omega)\big) \over \omega k_1^2 k_2^2}\ ,
\end{equation}
where
\begin{equation}
x(k\nd_1,k\nd_2,\omega)=1+{\omega^2-2(k\nd_1+k\nd_2)\,\omega\over 2k\nd_1 k\nd_2}\ .
\end{equation}
The constraints over the $k\nd_1$ and $k\nd_2$ integrals are the following.  First, since $\omega=k\nd_1+k\nd_2+k\nd_3$, we must
have $k\nd_1+k\nd_2 \le \omega$.  Second, we must have $x\le 1$, which gives $k\nd_1 + k\nd_2 > \half\omega$.  Finally, we must have
$x \ge -1$, which gives $(2k\nd_1-\omega)(2k\nd_2-\omega)\ge 0$.  Putting this all together, we find that $k\nd_1$ and $k\nd_2$ are to
be integrated over the shaded triangle in Fig.~\ref{irange}.

It is convenient to define $k\nd_1\equiv \half\omega u$ and $k\nd_2\equiv \half\omega v$.  Then the vertices of the triangle in the $(u,v)$
plane are $(1,0)$, $(1,1)$, and $(0,1)$.   We then have
\begin{align}
&\frac{1}{m^2 + k_1^2 - (k\nd_2 + k\nd_3)^2 }\big[ \frac{k_1^2 (N-1)}{m^2 + k_1^2 - (k\nd_2 + k\nd_3)^2 } - \frac{k_1^2 + k_2^2 - k_3^2}{m^2+ k_2^2 - (k\nd_1 + k\nd_3)^2 } \big] \bvph \nonumber\\
&\quad=\frac{\omega^2/4}{m^2-\omega^2 (1-u)} \big[ \frac{u^2(N-1)}{m^2-\omega^2 (1-u) } + \frac{2(uv-2u-2v+2)}{m^2-\omega^2(1-v) }\big]\bvph
\end{align}
as well as
\begin{equation}
x=1+{2\over uv}\,\big(1-u-v\big)\ .
\end{equation}

We obtain, in $d=3$ space dimensions, the low-frequency behavior $(\omega\ll m)$
\begin{align}
\sigma\nd_g(\omega) &= \frac{g m^4\,\omega^3}{2^{9} 3N\pi^3}\int\limits_0^1\!\!du\!\int\limits_{1-u}^1\!\!\!dv\ 
\frac{1}{m^2-\omega^2 (1-u) } \bigg[ \frac{u^2(N-1)}{m^2-\omega^2 (1-u) } \nonumber\\
&\hskip1.5in+\frac{2(uv-2u-2v+2)}{m^2-\omega^2 (1-v) }\bigg]\\
&=\frac{g m^3}{3\pi^2 2^{9}}\left\{ \frac{N-2}{N}\left(\frac{\omega^3}{4m^3}+\frac{\omega^5}{10m^5}\right) +
\frac{9N-16}{N}\frac{\omega^7}{180 m^7} +\ldots\right\}.\nonumber
\end{align}
For $d=2$ we find
\begin{align}
\sigma\nd_g(\omega)&= \frac{g m^4 \omega}{2^8N\pi^2} \int\limits_0^1\!\!du\!\int\limits_{1-u}^1\!\!\!dv\ 
\frac{1}{\sqrt{(u+v-1)(1-u)(1-v)}} \frac{1}{m^2-\omega^2 (1-u)}\times\nonumber \\
&\hskip 1.0in \times  \bigg[ \frac{u^2(N-1)}{m^2-\omega^2 (1-u) } + \frac{2(uv-2u-2v+2)}{m^2-\omega^2 (1-v) }\bigg]\\
&=\frac{g m}{2^{8}\pi}\left\{\frac{N-2}{N}\left(\frac{16\omega}{15m}+\frac{32\omega^3}{105 m^3}\right) +
\frac{3N-5}{N}\frac{16 \omega^5}{315 m^5}  + \ldots \right\}.\nonumber
\end{align}

Remarkably, in the case $N=2$, the conductivity vanishes for small powers of $\omega/m$ for both $d=2$ and $d=3$, leading to a strong pseudogap behavior in the optical conductivity.

\section{Layered lattice bosons with Coulomb interactions}
\label{app:LLB}

The  Lagrangian density for layered charged bosons is [see Ref. \cite{MA}, Eqs. (7-9)]
\begin{align}
\CL\nd_\RE&=\frac{1}{2gQ^2}(\bnabla{\dot\pi})^2 + \frac{1}{2g}{\dot\sigma}^2 + \frac{1}{2g} C\nd_{ij} \left(\nabla\nd_i\sigma\,\nabla\nd_j\sigma
+\nabla\nd_i\pi\,\nabla\nd_j\pi\right)\nonumber\\
&\qquad\qquad\qquad + \CL\nd_{\rm int},
\end{align}
where $Q^2=16 \pi e^2 /\epsilon$ is the effective electric charge of the bosons and $C=\textsf{diag}(1,1,\alpha)$, with $\alpha^2 \ll 1$
the ratio of $c$-axis and  in-plane superfluid stiffness.  The bare propagators are now
\bea
G_{\sigma\sigma}(\Bk) &=& {g \over   \omega_n^2 +  \Bk_{ab}^2+\alpha^2 k_c^2 +1 }\\
G_{\pi\pi}(\Bk) &=&  {g Q^2 \over \Bk^2 ( \omega_n^2 +\omega_p^2 f^2_p (\Bk)  ) }     \\
f_p(\Bk) &=&   \sqrt{   \Bk_{ab}^2   + \alpha^2_p k_c^2 \over k^2  } 
\eea
Here $\omega_{p}$ and $\alpha_p \omega_p$ are the in-plane and $c$-axis plasma frequencies given by
\beq
\omega_{p}^2 = {Q^2 \over a_c},~~~\alpha_p^2= { \alpha^2  a_c  \over a_{ab}^2 } \ ,
\eeq
where $a_i$ are the lattice constants.

We evaluate the $\CO(g^0)$ (one-loop) conductivity:
\begin{align}
K_{\mu\nu}(i\omega_m)&={1\over\beta}\sum_{\nu\nd_m}\int\!\!{d^d\!k\over (2\pi)^d}  4 k^\mu k^\nu  G_{\sigma\sigma}(i\nu_n,\Bk_{ab},k_c)\times\nonumber\\
&\quad\qquad\times
G_{\pi\pi}(i\omega_m+ i\nu_n,\Bk_{ab},k_c)
\end{align}
We set $\mu=\nu=x$ and $q=(i\omega\nd_n,\Bq=0)$.  We then have
\begin{align}
K_{xx}&=\frac{ 2 Q^2}{\beta}\sum_{\nu\nd_m} \!\!
\int \frac{d^2k_{ab}}{(2\pi)^2 k^2 }\int\frac{dk_{c}}{2\pi} \frac{\Bk_{ab}^2}{\Bk_{ab}^2+\alpha k_c^2 +1-(i\nu\nd_m)^2} \times\nonumber\\
&\qquad\qquad\times\frac{1}{\omega_p(\Bk)^2 - (i\nu\nd_m+i\omega\nd_n)^2}\\
&=\frac{Q^2}{2 \beta \pi^2  }\sum_{\nu\nd_m} \int  dk_{ab} \int dk_{c} \frac{ k_{ab}^3}{k^2} \frac{1}{(i\nu\nd_m+a\nd_k)(i\nu\nd_m-a\nd_k)}\times\nonumber\\
&\qquad\qquad\times\frac{1}{(i\nu\nd_m+i\omega\nd_n +b\nd_k)(i\nu\nd_m+i\omega\nd_n-b\nd_k) }\ , 
\end{align}
where
\begin{equation}
a\nd_\Bk=\sqrt{ \Bk_{ab}^2+\alpha k_c^2 +1},\qquad b\nd_\Bk= \omega_p f_p (\Bk)
\end{equation}
This is of similar form to Eq.~(\ref{eq:g0K}), but with a different choice of $a\nd_\Bk$ and $b\nd_\Bk$.  We can then use Eqs.~(\ref{eq:Fres}) and (\ref{eq:FT0}) to obtain
the conductivity for $\omega>0$,
\beq
\sigma(\omega) =  \frac{ Q^2}{8 \pi \omega  }   \int  dk_{ab} \int dk_{c}   {k_{ab}^3 \over a_k b_k k^2} \delta \left(\omega-a_k-b_k \right)
\eeq
Define $k_{ab}=k\cos\theta$, and anisotropy functions
\bea
f(\theta)&=& \sqrt{\cos^2\theta+\alpha^2 \sin^2\theta}\nonumber\\
f_p(\theta) &=& \sqrt{\cos^2\theta+\alpha^2_p \sin^2\theta}
\eea
such that
\beq
a\nd_\Bk=\sqrt{ 1+k^2 f^2(\theta)},\qquad b\nd_\Bk=\omega_{p} f_p(\theta) 
\eeq

The conductivity integrals can be evaluated  in cylindrical coordinates,  
\bea 
\sigma 
&=&\frac{ Q^2}{2\pi \omega\omega_{p} }  \int_0^\infty dk\, k^2 \int_0^{\pi/2} d\theta \frac{ \cos^3\theta}{\sqrt{1+k^2 f^2(\theta)}  f_p(\theta)}\times\nonumber \\
&\,&\qquad \times\delta\left(\omega- \sqrt{1+k^2 f^2(\theta)}- \omega_{p} f_p(\theta) \right)
 \eea
 Changing the argument of the delta function from $\omega$ to  $k^2$, provided $ (\omega-\omega_p f_p(\theta)) \ge 1$, yields
\bea
k_{\theta,\omega}^2 &\equiv &{ (\omega-\omega_p f_p(\theta))^2 -1 \over f^2(\theta)}
\eea
Thus,
\bea
\delta(\omega-\omega_{k^2})&=& \delta( k^2 - k^2_\omega) {dk_{\theta,\omega}^2 \over d\omega}\\
&=&2 \delta( k^2 - k^2_\omega) {  \sqrt{1+k^2 f^2(\theta)} \over f^2(\theta)}. 
\eea
This yields a simplified expression,
\begin{align}
 \sigma(\omega)&=\frac{Q^2}{\pi \omega\omega_{p}}  \int_0^{\pi/2} d\theta  { \cos^3\theta \over f_p(\theta) f^{3}(\theta)}  ~ \sqrt{(\omega-\omega_p f_p(\theta))^2-1}\,\times\nonumber\\
&\qquad\times\Theta\left( \omega-\omega_p f_p(\theta)  -1\right)
\end{align}
This expression is plotted in Fig.~\ref{sigmafig} for the parameters $\omega_p=10$, $\omega_c=0.1$, and $\alpha=10^{-2}$.



\begin{thebibliography}{99}
\bibitem{relativistic} Here we focus on relativistic theories and do not consider Galilean invariant superfluids or $O(3)$ ferromagnets, which admix the Goldstone and amplitude fluctuations.
\end{thebibliography}
\end{document}